\documentclass[12pt]{article}
\usepackage{amsmath}
\usepackage{graphicx,psfrag,epsf}
\usepackage{enumerate}
\usepackage{natbib}
\usepackage{url} 

\newcommand{\blind}{0}

\usepackage{xr}

\usepackage{color,enumitem,float,booktabs,makecell,amsthm,amssymb,amsfonts,dsfont,hyperref,multirow,booktabs,mathrsfs}
\hypersetup{
     colorlinks   = true,
  linkcolor=black,
            urlcolor=black,
            citecolor=black
}
\usepackage[font={stretch=1.0}]{caption} 

\def\lbk{\left\{}
\def\rbk{\right\}}
\def\lak{\left|}
\def\rak{\right|}
\def\lmk{\left[}
\def\rmk{\right]}
\def\lsk{\left(}
\def\rsk{\right)}
\def\bml{\boldsymbol{1}}
\def\bmo{\boldsymbol{0}}
\def\bmy{\boldsymbol{y}}

\def\bmbeta{\boldsymbol{\beta}}

\def\bmx{\boldsymbol{x}}
\def\bmb{\boldsymbol{b}}
\def\bms{\boldsymbol{s}}
\def\bmB{\boldsymbol{B}}
\def\bmA{\boldsymbol{A}}
\def\bmX{\boldsymbol{X}}
\def\bmM{\boldsymbol{M}}
\def\bmQ{\boldsymbol{Q}}

\def\bmS{\boldsymbol{S}}
\def\bmW{\boldsymbol{W}}
\def\bmI{\boldsymbol{I}}
\def\bmG{\boldsymbol{G}}
\def\bmUpsilon{\boldsymbol{\Upsilon}}
\def\bmXi{\boldsymbol{\Xi}}
\def\bmSigma{\boldsymbol{\Sigma}}

\def\bmtheta{\boldsymbol{\theta}}

\def\bmalpha{\boldsymbol{\alpha}}
\def\laak{\left\|}
\def\raak{\right\|}
\def\diag{{\rm diag}}
\def\rmvec{{\rm vec}}
\def\Cov{{\rm Cov}}

\def\rmP{{\rm P}}
\def\nn{\nonumber}
\def\bmV{\boldsymbol{V}}
\def\bmU{\boldsymbol{U}}

\def\bmOmega{\boldsymbol{\Omega}}

\def\tr{{\rm tr}}
\def\rmE{{\rm E}}
\def\rmS{{\rm S}}
\def\rmM{{\rm M}}

\def\rmd{{\rm d}}

\def\mF{\mathcal{F}}

\def\be{\begin{equation}}
\def\ee{\end{equation}} 
\def\ben{\begin{equation*}}
\def\een{\end{equation*}}
\def\bea{\begin{eqnarray}}
\def\eea{\end{eqnarray}}
\def\bda{\begin{eqnarray*}}
\def\eda{\end{eqnarray*}}
\def\beal{\begin{aligned}}
\def\eeal{\end{aligned}} 
\def\bmvarepsilon{\boldsymbol{\varepsilon}}

\newtheorem{tm}{Theorem}

\newtheorem{as}{Assumption}

\newtheorem*{as:4'}{Assumption 4$'$}

\numberwithin{equation}{section}

\newcommand{\lrnormmm}[1]{{\left\vert\kern-0.25ex\left\vert\kern-0.25ex\left\vert #1 \right\vert\kern-0.25ex\right\vert\kern-0.25ex\right\vert}}
\newcommand{\normmm}[1]{{\vert\kern-0.25ex\vert\kern-0.25ex\vert #1 \vert\kern-0.25ex\vert\kern-0.25ex\vert}}

\usepackage{cleveref}

\begin{document}

\def\spacingset#1{\renewcommand{\baselinestretch}%
{#1}\small\normalsize} \spacingset{1}


\if0\blind
{\spacingset{1.5} 
  \title{\bf Quasi-Score Matching Estimation for Spatial Autoregressive Model with Random  Weights Matrix and Regressors}
 \author{Xuan Liang and Tao Zou\footnote{Corresponding author, and all correspondence concerning the article should be directed to Tao Zou at \texttt{tao.zou@anu.edu.au}. }
    \\
    \textit{\normalsize Research School of Finance, Actuarial Studies and Statistics, }\\
    \textit{\normalsize The Australian National University, Canberra, ACT 2600, Australia}
    }
    \date{}
  \maketitle
} \fi

\if1\blind
{
  \bigskip
  \bigskip
  \bigskip
  \begin{center}
    {\LARGE\bf Title}
\end{center}
  \medskip
} \fi

\begin{abstract}

With the rapid advancements in technology for data collection, the application of the spatial autoregressive (SAR) model has become increasingly prevalent in real-world analysis, particularly when dealing with large datasets. However, the commonly used quasi-maximum likelihood estimation (QMLE) for the SAR model is not computationally scalable to handle the  data with a large size.  In addition, when establishing the asymptotic properties of the parameter estimators of the SAR model, both weights matrix and regressors are assumed to be nonstochastic in classical spatial econometrics, which is perhaps not realistic in real applications.  Motivated by the machine learning literature, this paper proposes quasi-score matching estimation for the SAR model. This new estimation approach is developed based on the likelihood,  but significantly reduces the computational complexity of the QMLE. The asymptotic properties of parameter estimators under the random weights matrix and regressors are established, which provides a new theoretical framework for the asymptotic inference of the SAR-type models. The usefulness of the quasi-score matching estimation and its asymptotic inference is illustrated via extensive simulation studies and a case study of an anti-conflict social network experiment for middle school students.

\end{abstract}

\noindent%
{\it Keywords:} Computational scalability, random weights matrix, random-X regression, score matching, spatial autoregressive model.
\vfill

\spacingset{1.8} 

\section{Introduction}
\label{sec:intro}


Spatial econometrics focuses on modeling spatial cross-sectional dependence, with the spatial autoregressive (SAR) model, introduced by \citet{cliff1973spatial}, being one of the most popular and widely used in the field (see, e.g., \citealp{anselin1988spatial} and  \citealp{lesage2009introduction}). Its clear structure and meaningful interpretation have made it a key tool in both research and real-world applications.










The SAR model is defined for spatial locations $i=1,\cdots,n$, where $y_i$ is the response variable and $\bmx_i$ is the associated $p$-dimensional covariate vector. Let $\bmy=(y_1,\cdots,y_n)^\top$ be the response vector and $\bmX=(\bmx_1,\cdots,\bmx_n)^\top$ be the $n\times p$ covariate matrix. The model is given by  
\be\label{eq:SAR}  
\bmy=\lambda \bmW \bmy+\bmX\bmbeta+\bmvarepsilon,\ee  
where $\bmvarepsilon=(\varepsilon_1,\cdots,\varepsilon_n)^\top$ is the error vector with mean $\bmo_n=(0,\cdots,0)^\top\in\mathbb{R}^n$ and covariance matrix $\sigma^2 \bmI_n$, $\bmI_n$ is the $n$-dimensional identity matrix, $\bmW=(w_{ij})_{n\times n}$ is the $n\times n$ spatial weights matrix, $\lambda\in\mathbb{R}$ measures spatial dependence, and $\bmbeta=(\beta_1,\cdots,\beta_p)^\top\in\mathbb{R}^p$ is the coefficient vector. The parameter of interest is $\bmtheta=(\lambda, \bmbeta^\top,\sigma^2)^\top$.







To estimate $\bmtheta$, the quasi-maximum likelihood estimation (QMLE) is widely used  in spatial econometrics \citep{lee2004asymptotic, lesage2009introduction}. With growing data sizes, SAR models are increasingly applied to large $n$-dimensional response vector $\bmy$ and weights matrix $\bmW$ \citep{huang2019least, huang2020two, zhu2020multivariate}, where $n$ can exceed thousands. However, computing the QMLE becomes challenging due to the $O(n^3)$ complexity of calculating the determinant $\det(\bmI_n-\lambda \bmW)$ in the quasi-log-likelihood; see the discussion in Section \ref{sec:estimation}. 
 To mitigate this,  empirical researchers often consider alternative methods such as the two-stage least squares (2SLS, \citealp{kelejian1998generalized}) estimation.  Another approach introduces the matrix exponential SAR model \citep{lesage2007matrix} in contrast to model (\ref{eq:SAR}),  to bypass determinant computation.


This article introduces a novel quasi-score matching estimation for the SAR model (\ref{eq:SAR}), inspired by score matching in machine learning (see, e.g., \citealp{hyvarinen2005estimation}), where score matching has been widely applied in the rapidly growing fields of  generative artificial intelligence and generative statistics (see, e.g., \citealp{cheng2024foundation}).
Unlike existing score matching methods for independent and identically distributed data, to our understanding, our approach is the first to address spatially dependent data. It estimates $\bmtheta$ without computing the determinant, thereby largely reducing computational cost for large $n$. 
 It is worth noting that although the quasi-score matching is established based on the likelihood (see Section \ref{sec:estimation}), it does not maximize the likelihood function. 
Accordingly, the quasi-score matching reduces computational cost while trading off some estimation efficiency compared to the QMLE; see Corollary 1 in Section S.4.1 of the supplementary material. 
However, based on the numerical estimation efficiencies demonstrated in various simulations in Section \ref{sec:simu}, we find that the quasi-score matching achieves efficiency close to the QMLE, especially for large $n$. Additionally, as shown in Table \ref{tb:data1} of our case study, its standard errors closely match those of the QMLE. 
Apart from the efficiency comparisons,  our simulation results in Section \ref{sec:simu} highlight its significant computational advantage, requiring only about 1/16,000 to 1/26,000 of the QMLE’s time for $n=$ 10,000. 
 Additional simulation studies in Section S.9.5 of the supplementary material compare our proposed quasi-score matching method with the generalized method of moments (GMM) estimation by \citet{lee2007gmm}, 
 the least squares estimation (LSE)  by \citet{huang2019least}, the 2SLS estimation  by \citet{kelejian1998generalized}, the best 2SLS estimation by \citet{lee2003best} and the one-step estimation by \citet{gupta2023efficient}. The results show that our method is computationally faster than all of these methods except 2SLS, particularly for large $n$, where 2SLS is only slightly faster but exhibits lower estimation efficiency.

In addition to the computational gains of the proposed quasi-score matching estimation for model (\ref{eq:SAR}), this paper also pioneers the study of its asymptotic properties when both the spatial weights matrix $\bmW$ and regressors $\bmX$ are random. 
Traditionally, $\bmW=(w_{ij})_{n\times n}$ is assumed nonstochastic, with weights $w_{ij}$  based on fixed geographical distances $d_{ij}$ or neighborhood structures  for $i,j=1,\cdots,n$ (see, e.g., \citealp{kelejian1998generalized}, \citealp{lee2004asymptotic}, \citealp{lesage2009introduction}). 
However, recent applications define the distance $d_{ij}$ using economic variables, such as GDP rankings (\citealp{yu2016strategic}), or set $\bmW$ as the normalized network adjacency matrix in social networks \citep{huang2019least,huang2020two,zhu2020multivariate}. 
Since economic variables and network adjacency matrices are observations or  realizations of random variables, assuming $\bmW$ to be stochastic is more realistic, aligning with stochastic network modeling frameworks (see, e.g., \citealp{bickel2009nonparametric} and \citealp{schweinberger2020exponential}). 

Aside from the weights matrix $\bmW$,  the models and numerical studies  in the  above literature also consider that the regressors $\bmx_i$ in  $\bmX=(\bmx_1,\cdots,\bmx_n)^\top$ are  observed or realized random vectors.  
However, for simplicity, some studies of the above literature assume that once observed, $\bmW$ and $\bmX$ are nonstochastic, with asymptotic results relying on this assumption (see, e.g., \citealp{kelejian1998generalized}, \citealp{kelejian2001asymptotic}, \citealp{lee2004asymptotic}, \citealp{huang2019least}, \citealp{huang2020two} and  \citealp{zhu2020multivariate}). 
This implicitly assumes $\bmx_i$s are independent across locations $i=1,\cdots, n$, as nonstochastic constants are mutually independent, and that $\bmX$ and $\bmW$ are also independent. 
However, in real applications, $\bmx_i$ is often spatially dependent in $i$ alongside $y_i$, making $\bmX$ potentially dependent on $\bmW$. 
As noted in Chapter 1.4 of \citet{wooldridge2002econometric}, assuming nonstochastic $\bmX$ may introduce limitations in econometric practice. 

Building on the above motivations, we propose a new theoretical framework to study the asymptotic properties of the proposed estimators under random $\bmW$ and $\bmX$, thereby largely broadening SAR model applications while making the classical nonstochastic setting a special case.   
To this end, we develop a new law of large numbers (LLN) and central limit theorem (CLT) for a general linear-quadratic form $\bms_n=(\bmvarepsilon^\top \bmA_1\bmvarepsilon,\cdots, \bmvarepsilon^\top \bmA_d\bmvarepsilon)^\top +\bmB^\top \bmvarepsilon$, where  without loss of generality,  the random matrices $\bmA_j\in\mathbb{R}^{n\times n}$ and $\bmB\in\mathbb{R}^{n\times d}$ for $j=1,\cdots,d$ are assumed to be measurable functions of  $\bmW$ and $\bmX$.   
Unlike \citet{kelejian2001asymptotic}, which considers only \( d = 1 \) with nonstochastic \( \bmA_1 \) and \( \bmB \), 
 as well as \citet{xu2015maximum} and \citet{wu2022applications}, which allow random \( \bmX \) but assume nonstochastic \( \bmW \), our framework accommodates randomness in both \( \bmW \) and \( \bmX \).
  It turns out that  
this new framework requires 
non-trivial conditions on $\bmvarepsilon$ beyond standard exogeneity,   
and the random $\bmX$ setting can relax the  condition of uniform boundedness in absolute value for the elements of $\bmX$ imposed in classical spatial econometrics (see, e.g., \citealp{kelejian1998generalized}, \citealp{kelejian2001asymptotic} and  \citealp{lee2004asymptotic}). More details are provided in Section \ref{sec:theory}.   
%
Notably, both the quasi-score matching estimation and our theoretical framework can be applied in various spatial econometrics models when their $\bmW$ and $\bmX$  are random (see, e.g., \citealp{lee2010estimation}, \citealp{zou2021network} and \citealp{wu2021inward}). 

The rest of this paper is organized as follows. Section \ref{sec:estimation} introduces the quasi-score matching estimation for the SAR model (\ref{eq:SAR}). Section \ref{sec:theory} establishes the new LLN and CLT  to study the consistency and  asymptotic normality of the quasi-score matching estimators.  Sections \ref{sec:simu} and \ref{sec:casestudy} provide simulation studies and a case study, respectively. Section \ref{sec:conclusion} gives concluding remarks. To save space,  additional simulation and case study results  are presented in Sections S.9 and S.10 of the supplementary material, respectively. 
All theoretical proofs, 
technical lemmas, and additional theoretical results including corollaries, are also relegated to the supplementary material.

\section{Quasi-Score Matching Estimation}

\label{sec:estimation}

In this section, we denote $\bmtheta_0=(\lambda_0,\bmbeta_0^\top,\sigma_0^2)^\top$ as the true parameter vector in model (\ref{eq:SAR}). Subsequently, the true SAR model becomes
\be\label{eq:SAR0}
\bmy=\lambda_0 \bmW \bmy+\bmX\bmbeta_0+\bmvarepsilon,
\ee
where the error vector $\bmvarepsilon$ has mean $\bmo_n$ and covariance matrix $\sigma_0^2 \bmI_n$.

In order to estimate  $\bmtheta_0$, the  QMLE is commonly considered. Specifically, the estimator maximizes the log-likelihood function based on a conditional Gaussian density of $\bmy$ given $\bmX$ and $\bmW$, which is 
\be\label{eq:cy|XW}
p_{\bmtheta}(\bmy|\bmX, \bmW)=\frac{1}{Z(\bmtheta|\bmW)}\exp\lmk -\frac{1}{2\sigma^2}\lbk \bmS(\lambda)\bmy-\bmX\bmbeta\rbk^\top\lbk \bmS(\lambda)\bmy-\bmX\bmbeta\rbk\rmk,
\ee
where $\bmS(\lambda)=\bmI_n-\lambda \bmW$, $Z(\bmtheta|\bmW)=(2\pi\sigma^2)^{n/2}  \lak \det \{\bmS(\lambda)\}\rak^{-1}$ is a normalizing constant such that $\int p_{\bmtheta}(\bmy|\bmX, \bmW)\rmd \bmy =1$, and $\det \{\bmS(\lambda)\}$ is the determinant of the $n\times n$ matrix $\bmS(\lambda)$. The QMLE can be obtained by maximizing the log-likelihood function
\be\label{eq:likelihood}
\ell_n(\bmtheta) =\ell_n(\lambda,\bmbeta,\sigma^2)=\log p_{\bmtheta}(\bmy|\bmX, \bmW).
\ee
Since the data $\bmy$ is not required to be Gaussian in the SAR model (\ref{eq:SAR0}), $\ell_n(\bmtheta)$ is called a quasi-log-likelihood. In real practice, however, the concentrated likelihood approach in \citet{lee2004asymptotic}  is adopted  instead of directly optimizing the quasi-log-likelihood function (\ref{eq:likelihood}). In particular, given $\lambda$, we first maximize $\ell_n(\lambda,\bmbeta,\sigma^2)$ with respect to $\bmbeta$ and $\sigma^2$, which yields the optimal $\bmbeta$ and $\sigma^2$ given below:
\be\label{eq:beta}
\tilde\bmbeta(\lambda)=(\bmX^\top \bmX)^{-1}\bmX^\top \bmS(\lambda)\bmy,\textrm{ and }
\ee
\be\label{eq:sigma2}
 \tilde\sigma^2(\lambda)=\frac{1}{n} \lbk \bmS(\lambda)\bmy-\bmX\tilde\bmbeta(\lambda)\rbk^\top\lbk \bmS(\lambda)\bmy-\bmX\tilde\bmbeta(\lambda)\rbk=\frac{1}{n} \bmy^\top\bmS(\lambda)^\top \bmM_{\bmX} \bmS(\lambda)\bmy,
\ee
where $\bmM_{\bmX}=\bmI_n-\bmX(\bmX^\top \bmX)^{-1}\bmX$. Then we substitute $\bmbeta$ and $\sigma^2$ in $\ell_n(\lambda,\bmbeta,\sigma^2)$ with $\tilde\bmbeta(\lambda)$ and $ \tilde\sigma^2(\lambda)$ to attain the concentrated log-likelihood function
\be\label{eq:clik}
 \ell_n\lsk \lambda, \tilde\bmbeta(\lambda), \tilde\sigma^2(\lambda)\rsk\propto \log|\det \{\bmS(\lambda)\}|
-\frac{n}{2}\log \lbk \bmy^\top\bmS(\lambda)^\top \bmM_{\bmX}\bmS(\lambda)\bmy \rbk\triangleq \ell_n^c(\lambda),
\ee
where we omit those constants that are irrelevant to $\lambda$ in $\ell_n\big( \lambda, \tilde\bmbeta(\lambda), \tilde\sigma^2(\lambda)\big)$.  Next, we maximize $\ell_n^c(\lambda)$ with respect to $\lambda$, substitute $\lambda$ in (\ref{eq:beta}) and (\ref{eq:sigma2}) with the optimal $\lambda$, and obtain the QMLE $\tilde\bmtheta=(\tilde\lambda,\tilde\bmbeta^\top,\tilde\sigma^2)^\top$, where
\be
\label{eq:QMLE}
 \tilde\lambda=\arg\underset{\lambda}{\max}\, \ell_n^c(\lambda), \quad\tilde\bmbeta=\tilde\bmbeta(\tilde\lambda) \quad\text{ and }\quad\tilde \sigma^2= \tilde\sigma^2(\tilde\lambda).
\ee

According to the above approach, one can see that  the calculation of the QMLE $\tilde\bmtheta=(\tilde\lambda,\tilde\bmbeta^\top,\tilde\sigma^2)^\top$ involves computing $\det\{\bmS(\lambda)\}$, the determinant of the $n\times n$ matrix $\bmS(\lambda)$, which has a computational complexity of $O(n^3)$. As a consequence, the QMLE is not computationally scalable for a large $n$.  This motivates us to propose an alternative approach to reduce the computational complexity of estimation. It is worth noting that the determinant $\det\{\bmS(\lambda)\}$ comes from the normalizing constant $Z(\bmtheta|\bmW)$ in the conditional density (\ref{eq:cy|XW}).  If we can avoid 
computing the normalizing constant $Z(\bmtheta|\bmW)$ in the estimation procedure, the computational complexity of  the QMLE will be reduced. To this end,   we adapt the score matching approach in the machine learning literature (see, e.g., \citealp{hyvarinen2005estimation}), which is originally proposed to remove the normalizing constant  in estimation.

In what follows, we demonstrate three important differences between our proposed score matching approach in this paper and the original score matching in \citet{hyvarinen2005estimation},  noting that a recent literature \citet{cheng2024foundation} proposed the maximum GIC estimator, which is equivalent to the original score matching estimator in \citet{hyvarinen2005estimation}.
(I) Based on our understanding, this article  is the first to propose a score matching approach for spatially dependent data $y_1,\cdots,y_n$ which follow the SAR model (\ref{eq:SAR0}), and the asymptotic results are established under $n\to\infty$; see Theorems \ref{tm:SM} -- \ref{tm:I} below. In contrast, the original score matching is proposed for $T$ independent and identical replications of $\bmy=(y_1,\cdots,y_n)^\top$, say $\bmy^{(1)},\cdots,\bmy^{(T)}$, where the asymptotic results of \citet{hyvarinen2005estimation} and \citet{cheng2024foundation} are based on fixed $n$ and $T\to\infty$. 
(II)   \citet{hyvarinen2005estimation} and \citet{cheng2024foundation} did not consider the regression setting. 
Therefore, their approach is formulated on the basis of the density of $\bmy$, say $p_{\bmtheta_0}(\bmy)$. On the contrary, the SAR model (\ref{eq:SAR0}) is under a regression setting and hence we build up our score matching approach via the conditional density $p_{\bmtheta}(\bmy|\bmX, \bmW)$ in (\ref{eq:cy|XW}). (III) The original formulation of score matching  assumes that $\bmy^{(t)}$ for $t=1,\cdots,T$ have a true density of $p_{\bmtheta_0}(\bmy)$.  However, following the idea of the QMLE, we do not require  data $\bmy$ to be Gaussian, i.e., $\bmy$ only needs to satisfy   the SAR model (\ref{eq:SAR0}) but does not necessarily follow the Gaussian density  $p_{\bmtheta_0}(\bmy|\bmX, \bmW)$. 
  Hence, we name our proposed approach quasi-score matching estimation.  In fact, the proposed  quasi-score matching is also developed based on the  density  $p_{\bmtheta}(\bmy|\bmX, \bmW)$  in (\ref{eq:cy|XW}), or equivalent, the likelihood, with details provided below.

Let 
\be\label{eq:qt}
q_{\bmtheta}(\bmy|\bmX, \bmW)=\exp\lmk -\frac{1}{2\sigma^2}\lbk\bmS(\lambda)\bmy-\bmX\bmbeta\rbk^\top\lbk\bmS(\lambda)\bmy-\bmX\bmbeta\rbk\rmk,\ee
and then we have
$
p_{\bmtheta}(\bmy|\bmX, \bmW)=q_{\bmtheta}(\bmy|\bmX, \bmW)/{Z(\bmtheta|\bmW)}.
$
To propose the quasi-score matching estimation, we define a squared distance by
\bea
J(\bmtheta)&=&\frac{1}{2}\int \laak   \frac{\partial }{\partial \bmy}\log p_{\bmtheta}(\bmy|\bmX, \bmW)-\frac{\partial }{\partial \bmy}\log p_{\bmtheta_0}(\bmy|\bmX, \bmW)\raak_2^2 p_{\bmtheta_0}(\bmy|\bmX, \bmW)\rmd \bmy\nn\\
&=&\frac{1}{2}\int \laak \frac{\partial }{\partial \bmy}\log q_{\bmtheta}(\bmy|\bmX, \bmW)-\frac{\partial }{\partial \bmy}\log q_{\bmtheta_0}(\bmy|\bmX, \bmW)  \raak_2^2 p_{\bmtheta_0}(\bmy|\bmX, \bmW)\rmd \bmy,\label{eq:Jt}
\eea
which measures the distance between  $\partial \log p_{\bmtheta}(\bmy|\bmX, \bmW)/{\partial \bmy}=(\partial \log p_{\bmtheta}(\bmy|\bmX, \bmW)/{\partial y_1},\cdots,$ $\partial \log p_{\bmtheta}(\bmy|\bmX, \bmW)/{\partial y_n})^\top$ and $\partial \log p_{\bmtheta_0}(\bmy|\bmX, \bmW)/{\partial \bmy}$, where $\|\cdot\|_2$ is the Euclidean norm. It is worth noting that the machine learning literature refers $\partial \log p_{\bmtheta}(\bmy|\bmX, \bmW)/{\partial \bmy}$ as a ``score'' function (see, e.g., \citealp{hyvarinen2005estimation}), which is  different from the classical score function $\partial \log p_{\bmtheta}(\bmy|\bmX, \bmW)/{\partial \bmtheta}$ considered in econometrics. In spite of this, we still follow the machine literature and use the terminology ``score matching" in this paper. 

 Motivated by the score matching literature, there are two motivations for considering the squared distance $J(\bmtheta)$. First, the normalizing constant $Z(\bmtheta|\bmW)$ in (\ref{eq:cy|XW}) disappears within the norm $\|\cdot\|_2$; see (\ref{eq:Jt}). This will eliminate the need to compute $Z(\bmtheta|\bmW)$ in our quasi-score matching estimation introduced later. Second, $J(\bmtheta)$ in (\ref{eq:Jt}) achieves its minimum, $J(\bmtheta_0) = 0$, at $\bmtheta = \bmtheta_0$. Consequently, minimizing $J(\bmtheta)$ yields the true parameter vector $\bmtheta_0$.

Next,  Lemma 1 in the supplementary material demonstrates that (\ref{eq:Jt}) can be transformed as follows:
\bea
J(\bmtheta)&=&\int \sum_{i=1}^n \lmk\frac{\partial^2}{\partial y_i^2}\log q_{\bmtheta}(\bmy|\bmX, \bmW)+\frac{1}{2}\lbk \frac{\partial}{\partial y_i}\log q_{\bmtheta}(\bmy|\bmX, \bmW)\rbk^2\rmk p_{\bmtheta_0}(\bmy|\bmX, \bmW)\rmd \bmy\nn\\&&+\textrm{ constant},\label{eq:jtc}
\eea
where the constant in (\ref{eq:jtc}) is irrelevant to $\bmtheta$. This result is obtained via integration by parts; see the proof of Lemma 1 in the supplementary material.  Since $\bmtheta_0$ minimizes $J(\bmtheta)$, if we ignore the constant in (\ref{eq:jtc}),
we obtain that  $\bmtheta_0$  should also minimize
\[
D(\bmtheta) =\int \sum_{i=1}^n \lmk\frac{\partial^2}{\partial y_i^2}\log q_{\bmtheta}(\bmy|\bmX, \bmW)+\frac{1}{2}\lbk \frac{\partial}{\partial y_i}\log q_{\bmtheta}(\bmy|\bmX, \bmW)\rbk^2\rmk p_{\bmtheta_0}(\bmy|\bmX, \bmW)\rmd \bmy.
\]
Let 
\be\label{eq:dt}
D_n(\bmtheta)=\sum_{i=1}^n \lmk\frac{\partial^2}{\partial y_i^2}\log q_{\bmtheta}(\bmy|\bmX, \bmW)+\frac{1}{2}\lbk \frac{\partial}{\partial y_i}\log q_{\bmtheta}(\bmy|\bmX, \bmW)\rbk^2\rmk.
\ee
We subsequently have $\rmE_{\bmy\sim p_{\bmtheta_0}(\bmy|\bmX, \bmW)}\{D_n(\bmtheta)\}=D(\bmtheta)$, where $\rmE_{\bmy\sim p_{\bmtheta_0}(\bmy|\bmX, \bmW)}\{\cdot\}$ indicates the expectation taken over $\bmy$ with $\bmy$  following the conditional density $p_{\bmtheta_0}(\bmy|\bmX, \bmW)$. As a consequence, minimizing the objective function $D_n(\bmtheta)$ leads to an $M$-estimator (\citealp{van1998asymptotic}).  This estimator is then the quasi-score matching estimator proposed in this paper.  Since the normalizing constant $Z(\bmtheta|\bmW)=(2\pi\sigma^2)^{n/2}  \lak\det\{\bmS(\lambda)\}\rak^{-1}$  also disappears in the quasi-score matching objective function $D_n(\bmtheta)$ in (\ref{eq:dt}),   obtaining the quasi-score matching estimator does not require the computation of the $n\times n$ matrix $\bmS(\lambda)$'s determinant, and hence reduces the computational complexity. 

To further demonstrate the computational complexity of the quasi-score matching estimation, we derive the closed-form of $D_n(\bmtheta)$ based on (\ref{eq:qt}) and (\ref{eq:dt}), which is given below:
\bea
&&D_n(\bmtheta)=D_n(\lambda,\bmbeta,\sigma^2)\nn\\&=&-\frac{1}{\sigma^2}\tr\lbk\bmS(\lambda)^\top \bmS(\lambda)\rbk+\frac{1}{2\sigma^4}\lbk\bmS(\lambda)\bmy-\bmX\bmbeta\rbk^\top\bmS(\lambda)\bmS(\lambda)^\top \lbk\bmS(\lambda)\bmy-\bmX\bmbeta\rbk. \label{eq:obj}
\eea
Similar to the QMLE, we adopt the concentrated approach when minimizing the objective function $D_n(\bmtheta)$, 
and obtain the quasi-score matching estimator.  In particular, we obtain the first-order condition (FOC) of minimizing $D_n(\bmtheta)$, which is ${\partial D_n(\bmtheta)}/{\partial \bmtheta}=\big( {\partial D_n(\bmtheta)}/{\partial \lambda},{\partial D_n(\bmtheta)}/{\partial \bmbeta^\top},{\partial D_n(\bmtheta)}/{\partial \sigma^2}\big)^\top=\bmo_{p+2}$, where the closed form of ${\partial D_n(\bmtheta)}/{\partial \bmtheta}$ is given in Lemma 11 of the supplementary material. Solving $\big( {\partial D_n(\bmtheta)}/{\partial \bmbeta^\top},{\partial D_n(\bmtheta)}/{\partial \sigma^2}\big)^\top=\bmo_{p+1}$ for a given $\lambda$, yields  
 the optimal  $\bmbeta$ and $\sigma^2$:
\be\label{eq:smbeta}
\hat\bmbeta(\lambda)=\lbk \bmX^\top \bmS(\lambda)\bmS(\lambda)^\top \bmX\rbk^{-1}\bmX^\top \bmS(\lambda)\bmS(\lambda)^\top\bmS(\lambda)\bmy,\textrm{ and }
\ee
\be\label{eq:smsigma2}
 \hat\sigma^2(\lambda)=\frac{1}{\tr\lbk\bmS(\lambda)^\top \bmS(\lambda)\rbk} \bmy^\top\bmS(\lambda)^\top\bmS(\lambda) \bmQ_{\bmX}(\lambda) \bmS(\lambda)^\top\bmS(\lambda)\bmy,
\ee
where $\bmQ_{\bmX}(\lambda)=\bmI_n-\bmS(\lambda)^\top \bmX\lbk \bmX^\top \bmS(\lambda)\bmS(\lambda)^\top \bmX\rbk^{-1}\bmX^\top \bmS(\lambda)$. We subsequently substitute $\bmbeta$ and $\sigma^2$ in $(\ref{eq:obj})$ with $\hat\bmbeta(\lambda)$ and $ \hat\sigma^2(\lambda)$,  to attain the concentrated $D_n(\lambda,\bmbeta,\sigma^2)$, which is
\ben
D_n\lsk \lambda, \hat\bmbeta(\lambda), \hat\sigma^2(\lambda)\rsk= -\frac{\tr^2\lbk\bmS(\lambda)^\top \bmS(\lambda)\rbk}{2 \bmy^\top\bmS(\lambda)^\top\bmS(\lambda) \bmQ_{\bmX}(\lambda) \bmS(\lambda)^\top\bmS(\lambda)\bmy}\triangleq D_n^c(\lambda).
\een
Then the quasi-score matching estimators of $\lambda$, $\bmbeta$ and $\sigma^2$ are obtained via 
\be
\label{eq:sm}
\hat\lambda=\arg \underset{\lambda}{\min}\, D_n^c(\lambda),  \quad\hat\bmbeta=\hat\bmbeta(\hat\lambda)\quad\text{ and }\quad \hat \sigma^2= \hat\sigma^2(\hat\lambda).
\ee 
 In Remark 4 of Section S.7 in the supplementary material, we demonstrate the numerical visualization of  the function $D^c_n(\lambda)$  based on the simulation settings considered in Section \ref{sec:simu}, showing that minimizing $D_n^c(\lambda)$  yields a unique solution numerically.

 Based on the above concentrated procedure, for a given $\lambda$, the quasi-score matching estimator $\hat\bmbeta(\lambda)$ in (\ref{eq:smbeta}) is a weighted least squares estimator, whereas the QMLE $\tilde\bmbeta(\lambda)$ in (\ref{eq:beta}) is an ordinary least squares estimator. Introducing the unnecessary weight $\bmS(\lambda)\bmS(\lambda)^\top$ in $\hat\bmbeta(\lambda)$ naturally reduces the asymptotic efficiency of the quasi-score matching estimator; also see Section S.4.1 and Corollary 1 of the supplementary material, which theoretically show that the QMLE $\tilde\bmtheta$ is asymptotically more efficient than the quasi-score matching estimator $\hat\bmtheta=(\hat\lambda,\hat\bmbeta^\top,\hat\sigma^2)^\top$. To improve the asymptotic efficiency, we can substitute $\lambda$ in QMLE (\ref{eq:beta}) and (\ref{eq:sigma2})  with the quasi-score matching estimator $\hat\lambda$, yielding an improved estimator introduced in Section \ref{subsec:improve}.

Apart from trading off some estimation efficiency compared to the QMLE, one can see that the computational complexity of obtaining the quasi-score matching estimator (\ref{eq:sm}) is of order $O(n^2)$ for  a large $n$; this is because the computation of  $\det\{\bmS(\lambda)\}$ is not needed, and the computational complexity is mainly determined by $n$-dimensional matrices' multiplications. If we consider that the weights matrix $\bmW$ has at most $c$ non-zeros entries per row and per column (see, e.g.,  some $\bmW$ considered in \citealp{kelejian1998generalized} and \citealp{yu2016strategic} given each location having at most $c$ neighbors, and some normalized network adjacency matrices $\bmW$  considered in \citealp{huang2019least}, \citealp{huang2020two} and \citealp{zhu2020multivariate}), the computational complexity can be further reduced to $O(c n)$, which is linear to  $n$ for a finite $c$. Accordingly, the quasi-score matching estimation is more computationally scalable than the QMLE as $n$ is large.   It is worth noting that our proposed quasi-score matching estimator is also superior to the GMM estimator by \citet{lee2007gmm} and the one-step estimator by \citet{gupta2023efficient}  in terms of computation. Notably, both methods require computing $\bmS^{-1}(\lambda)$, which has a computational complexity of $O(n^3)$. Numerical comparisons with these and other existing methods are provided in Section S.9.5 of the supplementary material.

In addition to  analyzing the computational complexity, we next study the asymptotic properties of the quasi-score matching estimator $\hat\bmtheta=(\hat\lambda,\hat\bmbeta^\top,\hat\sigma^2)^\top$ for the SAR model (\ref{eq:SAR0}) with both random  weights matrix $\bmW$ and regressors $\bmX$.

\section{Theoretical Results}
\label{sec:theory}






To develop the asymptotic properties of the quasi-score matching estimator $\hat\bmtheta$, the main steps include establishing the CLT for $n^{-1/2}\partial D_n(\bmtheta_0)/\partial \bmtheta$ and  the LLN for $n^{-1}\partial^2 D_n(\bmtheta_0)/$ $(\partial \bmtheta\partial \bmtheta^\top)$. Denote $[\bmG]_s=(\bmG+\bmG^\top)/2$ as the symmetrization of any generic matrix $\bmG$, and $\bmo_{n_1\times n_2}$ as the $n_1\times n_2$ matrix of zeros.  By Lemma 11 of the supplementary material, we obtain  that 
\be\label{eq:score}\frac{\partial D_n(\bmtheta_0)}{\partial \bmtheta}=\begin{pmatrix}\bmvarepsilon^\top \bmA_1(\bmtheta_0) \bmvarepsilon\\\vdots\\\bmvarepsilon^\top \bmA_{p+2}(\bmtheta_0) \bmvarepsilon\end{pmatrix}+\bmB(\bmtheta_0)^\top\bmvarepsilon-\begin{pmatrix}\sigma_0^2\tr\{\bmA_1(\bmtheta_0)\}\\\vdots\\\sigma_0^2\tr\{\bmA_{p+2}(\bmtheta_0)\}
\end{pmatrix},
\ee where
$
\bmA_1(\bmtheta_0)=-\lbk \lmk \bmS(\lambda_0) \bmS(\lambda_0)^\top \bmW \bmS^{-1}(\lambda_0)\rmk_s
+ \lmk \bmS(\lambda_0)\bmW^\top \rmk_s\rbk/\sigma_0^4$, $
\bmA_2(\bmtheta_0)=\cdots=\bmA_{p+1}$ $(\bmtheta_0)=\bmo_{n\times n}$, $\bmA_{p+2}(\bmtheta_0)=-\bmS(\lambda_0)\bmS(\lambda_0)^\top/\sigma_0^6$,  and 
\be\label{eq:msym}
\bmB(\bmtheta_0)^\top=\begin{pmatrix}-\frac{1}{\sigma_0^4}\lbk \bmW\bmS^{-1}(\lambda_0) \bmX\bmbeta_0\rbk^\top \bmS(\lambda_0)\bmS(\lambda_0)^\top\\
-\frac{1}{\sigma_0^4}\bmX^\top \bmS(\lambda_0) \bmS(\lambda_0)^\top \\
\bmo_{1\times n}
\end{pmatrix}.
\ee
Note that the first two terms on the right hand side of (\ref{eq:score}) constitute a 
linear-quadratic form of $\bmvarepsilon$ with matrices $\bmA_j(\bmtheta_0)$ and $\bmB(\bmtheta_0)$ for $j=1,\cdots,(p+2)$, where $\bmA_j(\bmtheta_0)$ and $\bmB(\bmtheta_0)$ are measurable functions of $\bmW$ and $\bmX$. Meanwhile,  $\partial^2 D_n(\bmtheta_0)/(\partial \bmtheta\partial \bmtheta^\top)$ also contains a linear-quadratic form of $\bmvarepsilon$ with matrices that are functions of $\bmW$ and $\bmX$; see the closed form of  $\partial^2 D_n(\bmtheta_0)/(\partial \bmtheta\partial\bmtheta^\top)$ given in  (S.46) of the supplementary material. Therefore, the key to obtain the asymptotic properties of the quasi-score matching estimator is to establish the LLN and CLT for a general linear-quadratic form 
\be\label{eq:lq}
\bms_n=(\bmvarepsilon^\top \bmA_1\bmvarepsilon,\cdots, \bmvarepsilon^\top \bmA_d\bmvarepsilon)^\top +\bmB^\top \bmvarepsilon,\ee
where matrices $\bmA_j$ and $\bmB$ are measurable functions of  $\bmW$ and $\bmX$ for $j=1,\cdots, d$ and $1\leq d<\infty$.  Without loss of generality, we only need to consider $\bmA_j$ being symmetric in the linear-quadratic form of (\ref{eq:lq}); this is because that if $\bmA_j$ is not symmetric, the quadratic form $\bmvarepsilon^\top\bmA_j\bmvarepsilon\equiv \bmvarepsilon^\top[\bmA_j]_s\bmvarepsilon$, where the symmetrization $[\bmA_j]_s$ is always a symmetric matrix.

As discussed in Introduction, in real data analysis it is often the case that $\bmW$ and $\bmX$ are stochastic, which leads to matrices $\bmA_j$ and $\bmB$ in (\ref{eq:lq}) being random. In Section \ref{subsec:CLT}, we develop a new  LLN and a new CLT, which are suitable for the linear-quadratic form $\bms_n$ with such random matrices. It is worthing noting that the CLT developed in \citet{kelejian2001asymptotic} only considers $d=1$ and matrices $\bmA_1$ and $\bmB$ being nonstochastic. 
Hence, our theoretical results are more general, and can be applied to the models with random $\bmW$ and $\bmX$.   In addition, it turns out that  a non-trivial condition is needed for the error vector $\bmvarepsilon$ compared to the nonstochastic case, which is also discussed in Section \ref{subsec:CLT} below  and Section S.1 of the supplementary material.  Next, we apply the new LLN and CLT, and obtain the asymptotic properties of the quasi-score matching estimator in Section \ref{subsec:asyQS}. Lastly, we make use of  the QMLE (\ref{eq:beta}) and (\ref{eq:sigma2}), and the quasi-score matching estimator of $\lambda$, to obtain new estimators of $\bmbeta$ and $\sigma^2$ in Section \ref{subsec:improve}. We further demonstrate that doing this will remain the same computational complexity, but can improve the asymptotic efficiency  of $\bmbeta$ and $\sigma^2$ compared to the quasi-score matching estimation.

 \subsection{LLN and CLT for Linear-Quadratic Forms with Random Matrices}
 
\label{subsec:CLT}

To establish the LLN and CLT of the linear-quadratic form $\bms_n$ in (\ref{eq:lq}),  note that $\bmA_j$ and $\bmB$    are measurable functions of  random $\bmW$ and $\bmX$ for $j=1,\cdots,d$ mentioned below (\ref{eq:lq}). Hence, $\bmA_j$ and $\bmB$ are random and  $\mF_{\bmX,\bmW}$-measurable, where $\mF_{\bmX,\bmW}=\sigma\langle \bmX,\bmW\rangle$ is the $\sigma$-algebra generated by random matrices $\bmX$ and $\bmW$. 
In econometrics, it is common to impose the exogenous condition for random regressors $\bmX$ with respect to random errors 
$\varepsilon_i$. Adapting it to spatial econometrics with both random $\bmX$ and $\bmW$, we consider the exogenous condition given below:

\bigskip

\textbf{(C1)} $\rmE[\varepsilon_i|\mF_{\bmX,\bmW}]=\rmE[\varepsilon_i|\bmX,\bmW]=0$ for $i=1,\cdots,n$.

\bigskip

Next, we introduce a non-trivial condition for random errors $\varepsilon_i$. 

\bigskip

 \textbf{(C2)} Assume that $\varepsilon_i$ are conditional independent in $i$ given $\sigma$-algebra $\mF_{\bmX,\bmW}$ for $i=1,\cdots,n$.
 
 \bigskip

\noindent It is worth noting that the classical spatial econometrics (see, e.g., \citealp{lee2004asymptotic} and \citealp{lesage2009introduction}) often considers the more commonly assumed condition given below for random errors $\varepsilon_i$ under the  nonstochastic setting of $\bmW$ and $\bmX$: 

 \bigskip
 
 \textbf{(C2$'$)} Assume that $\varepsilon_i$ are independent in $i$ for $i=1,\cdots,n$.
 
 \bigskip

 \noindent However, independence condition (C2$'$) does not imply conditional independence condition (C2), and the opposite implication is also not valid (see \citealp{stoyanov2013counterexamples} for counterexamples). Further discussion on the necessity of the non-trivial Condition (C2) for establishing the  CLT of the linear-quadratic form with random matrices is presented in Section S.1 of the supplementary material.

After discussing the conditions for random errors $\varepsilon_i$ in Conditions (C1) -- (C2) above, 
we impose conditions for random matrices $\bmA_j$ and $\bmB$  in (\ref{eq:lq}) as follows.

\bigskip

\textbf{(C3)} Suppose that $\bmA_j=(a_{i_1i_2,j})_{n\times n}$ are $n\times n$ random symmetric matrices and $\bmB=(\bmb_1,\cdots,\bmb_d)=(b_{ij})_{n\times d}$  is an $n\times d$ random matrix for $i,i_1,i_2\in\{1,\cdots,n\}$ and $j=1,\cdots, d$, where  $\bmb_j$ is the $j$-th column of $\bmB$. Assume that $\bmA_j$, $j=1,\cdots, d$, and $\bmB$ are  all $\mF_{\bmX,\bmW}$-measurable. In addition, assume that (i) there exists $\eta_1>0$ such that $\rmE|\varepsilon_i|^{4+\eta_1}<\infty$; (ii) there exists $\eta_2\geq 8/\eta_1$ such that $\rmE|a_{i_1i_2,j}|^{2+\eta_2}<\infty$; and (iii) there exists $\eta_3\geq 4/(2+\eta_1)$ such that $\rmE|b_{ij}|^{2+\eta_3}<\infty$.

\bigskip

 We next introduce a stronger condition  for random matrices $\bmA_j$ and $\bmB$ below, which is used to obtain the new LLN for the linear-quadratic form (\ref{eq:lq}). 
Before introducing this condition, let $\|\cdot\|_\eta$ be the vector $\eta$-norm or the matrix $\eta$-norm for $1\leq \eta\leq \infty$.  Specifically, for any generic vector $\bmalpha=(\alpha_1,\cdots,\alpha_q)\in\mathbb{R}^q$, $\|\bmalpha\|_\eta=(\sum_{k=1}^q|\alpha_k|^\eta)^{1/\eta}$, and, for any generic matrix $\bmG\in\mathbb{R}^{m\times q}$, 
\be\label{eq:matrixnorm}
\laak \bmG\raak_\eta=\sup\lbk \frac{\laak \bmG\bmalpha\raak_\eta}{\laak \bmalpha\raak_\eta}:\bmalpha\in\mathbb{R}^q\textrm{ and }\bmalpha\neq \bmo_q\rbk.
\ee
If matrix $\bmG=(g_{k_1k_2})_{m\times q}$ is random, then, for $1\leq \eta\leq\infty$, denote $\|g_{k_1k_2}\|_{L^\eta}=(\rmE|g_{k_1k_2}|^\eta)^{1/\eta}$ as the $L^\eta$-norm of the random variable $g_{k_1k_2}$. Subsequently, let
\be\label{eq:Lnorm}
\laak \bmG\raak_{L^\eta}=\lsk \laak g_{k_1k_2}\raak_{L^\eta}\rsk_{m\times q}
\ee
and it is an $m\times q$ matrix consisting of the $L^\eta$-norm of $g_{k_1k_2}$. 
Moreover, denote
\be\label{eq:normmm}
\lrnormmm{\bmG}_{L^{\eta_1},\eta_2}=\big\| \laak \bmG\raak_{L^{\eta_1}}\big\|_{\eta_2}
\ee
for $\eta_1,\eta_2\in[1,\infty]$, where $\eta_1$ and $\eta_2$ correspond to the $L^{\eta_1}$-norm (\ref{eq:Lnorm}) and the matrix $\eta_2$-norm (\ref{eq:matrixnorm}), respectively.  Lastly, let $\rmvec(\bmG)$ be the vectorization for any generic matrix $\bmG$.

\bigskip

\textbf{(C4)} For $\eta_1$, $\eta_2$ and $\eta_3$ in (C3), assume that (i)  $\sup_i\rmE|\varepsilon_i|^{4+\eta_1}<\infty$; (ii)  $\sup_{n\geq 1}\normmm{\bmA_j}_{L^{2+\eta_2},1}$ $<\infty$ for $j=1,\cdots,d$; and (iii)  $\sup_{n\geq 1}n^{-1}\rmE\|\rmvec(\bmB)\|_{2+\eta_3}^{2+\eta_3}<\infty$.

\bigskip

Conditions (C3) and (C4) only impose some moment conditions on random matrices $\bmA_j$ and $\bmB$.  Compared to the literature, Assumption 2 in \citet{kelejian2001asymptotic}  is the nonstochastic version of these conditions for the  linear-quadratic form $\bms_n$ in (\ref{eq:lq}) with fixed $\bmA_j$ and $\bmB$, and $d=1$. However, since we consider matrices $\bmA_j$ and $\bmB$ being random, their $(2+\eta_2)$-th  and $(2+\eta_3)$-th order moments, respectively, need to  be coordinated with $\varepsilon_i$'s $(4+\eta_1)$-th order moments in a way of those demonstrated in Conditions (C3) -- (C4). 


Let $\bmUpsilon^{(2)}=\diag\{\mu_1^{(2)},\cdots,\mu_n^{(2)}\}$,  $\bmUpsilon^{(3)}=\diag\{\mu_1^{(3)},\cdots,$ $\mu_n^{(3)}\}$ and $\bmUpsilon^{(4)}=\diag\{\mu_1^{(4)}-3(\mu_1^{(2)})^2,\cdots,\mu_n^{(4)}-3(\mu_n^{(2)})^2\}$ with $\mu_i^{(s)}=\rmE[\varepsilon_i^s|\mF_{\bmX,\bmW}]$ for $s=2,3,4$,  where $\diag\{d_{1},\cdots,d_{n}\}$ denotes an $n\times n$ diagonal matrix with diagonals being $d_{1},\cdots,d_{n}$.    In addition, let $(g_{k_1k_2})_{m\times q}$ denote a generic $m\times q$ matrix with the $(k_1,k_2)$-th element being $g_{k_1k_2}$ for $k_1=1,\cdots, m$ and $k_2=1,\cdots,q$, and $\circ$  represent the Hadamard product of any two matrices with the same dimensions. The above notation and Conditions (C1)  -- (C4) yield the following theorem for the linear-quadratic form $\bms_n$ of  (\ref{eq:lq}).

\begin{tm}\label{tm:LLN} Under Conditions (C1) -- (C3), we have that (i)
\[
\rmE\lmk \bms_n|\mF_{\bmX,\bmW}\rmk=\lsk \tr(\bmUpsilon^{(2)}\bmA_{1}),\cdots,\tr(\bmUpsilon^{(2)}\bmA_{d})\rsk^\top,\textrm{ and }
\]
\bda
\Cov[\bms_n|\mF_{\bmX,\bmW}]&=&2\lsk \tr(\bmUpsilon^{(2)}\bmA_{j_1}\bmUpsilon^{(2)}\bmA_{j_2})\rsk_{d\times d}+\bmB^\top\bmUpsilon^{(2)}\bmB+\lsk \tr(\bmA_{j_1}\circ \bmUpsilon^{(4)}\circ \bmA_{j_2})\rsk_{d\times d}\\
&&+2\lmk \lsk\tr(\bmb_{j_1}\bml_n^\top\circ \bmUpsilon^{(3)}\circ \bmA_{j_2})\rsk_{d\times d}\rmk_s
\eda
where $\bml_n=(1,\cdots,1)^\top$ is the $n$-dimensional vector of ones; and (ii) (LLN)  if Condition (C4) is additionally assumed, we obtain 
\[
n^{-1/2-\eta}\lsk \bms_n-\rmE\lmk \bms_n|\mF_{\bmX,\bmW}\rmk\rsk\stackrel{\rmP}\longrightarrow \bmo_d,
\]
as $n\to\infty$ for any $\eta>0$.
\end{tm}

Theorem \ref{tm:LLN} (i) provides the closed forms of conditional mean and conditional covariance matrix of the linear-quadratic form $\bms_n$.  This is further utilized to show the LLN in Theorem  \ref{tm:LLN} (ii), and to obtain the closed form of $\Cov[\bms_n|\mF_{\bmX,\bmW}]$,  the latter of which is required in the CLT  for the linear-quadratic form $\bms_n$  in Theorem \ref{tm:CLT} given below.  It is worth noting that $\eta$ in Theorem \ref{tm:LLN} (ii) is a general positive constant. When we apply Theorem \ref{tm:LLN} (ii) in developing the theorems in the following subsections, we mainly use the result $n^{-1}\lsk \bms_n-\rmE\lmk \bms_n|\mF_{\bmX,\bmW}\rmk\rsk\stackrel{\rmP}\longrightarrow \bmo_d$ with $\eta=1/2$.

 In addition to  Conditions (C1) -- (C4) introduced earlier, the CLT of Theorem \ref{tm:CLT} requires an extra condition:

\bigskip

\textbf{(C5)} Assume that (C4) holds for $\eta_2>8/\eta_1$ and $\eta_3>4/(2+\eta_1)$. 

\bigskip

\noindent The above condition, together with Conditions (C1) -- (C4),  leads to the CLT for $\bms_n$.

\begin{tm} \label{tm:CLT}(CLT) Under Conditions (C1) -- (C5), if $n^{-1}\Cov[\bms_n|\mF_{\bmX,\bmW}]\stackrel{\rmP}\longrightarrow \bmSigma$ for a $d\times d$ finite and positive definite matrix $\bmSigma$, then
\[
n^{-1/2}\lsk \bms_n-\rmE[\bms_n|\mF_{\bmX,\bmW}]\rsk\stackrel{d}\longrightarrow N(\bmo_d,\bmSigma),
\]
where the closed form of $\Cov[\bms_n|\mF_{\bmX,\bmW}]$ is given in Theorem \ref{tm:LLN}.
\end{tm}

As mentioned in the beginning of this section, the CLT of Theorem \ref{tm:CLT} is developed to show the asymptotic normality of $n^{-1/2}\partial D_n(\bmtheta_0)/\partial \bmtheta$,  and the LLN of Theorem \ref{tm:LLN} (ii) is established to prove  that $n^{-1}\partial^2 D_n(\bmtheta_0)/(\partial \bmtheta\partial \bmtheta^\top)$ converges in probability. Both of these results are given in Lemma 11 of the supplementary material,  and they further lead to 
 the asymptotic normality of the quasi-score matching estimator $\hat\bmtheta$ in Theorem \ref{tm:SM} of the next subsection.   It is worth noting that  for the generality reason, both Theorems \ref{tm:LLN} and  \ref{tm:CLT} are developed under the heterogeneity of random errors $\varepsilon_i$. The classical SAR model (\ref{eq:SAR0}), on the other hand, assumes the  homoskedasticity of random errors $\varepsilon_i$, which is a special case. In addition,  Theorems \ref{tm:LLN} and  \ref{tm:CLT} can be applied in many other spatial econometrics  models with random $\bmW$ and $\bmX$,  as long as the linear-quadratic form is involved in deriving the asymptotic properties of  estimators   (see, e.g., \citealp{lee2010estimation},  \citealp{zou2021network} and \citealp{wu2021inward}).

\subsection{Asymptotic Results for Quasi-Score Matching Estimator}
\label{subsec:asyQS}

In this subsection, we apply Theorems \ref{tm:LLN} and  \ref{tm:CLT}  to establish the asymptotic normality of the proposed quasi-score matching estimator $\hat\bmtheta$ for the SAR model (\ref{eq:SAR0}) with both random  weights matrix $\bmW$ and regressors $\bmX$ in Theorem \ref{tm:SM} below.   We first lay out the technical assumptions required in this theorem.

\begin{as}\label{as:1}For model (\ref{eq:SAR0}), assume that the random error vector $\bmvarepsilon=(\varepsilon_1,\cdots,\varepsilon_n)^\top$ satisfies Conditions (C1), (C2) and (C4)-(i) in Section \ref{subsec:CLT},  where the condition for $\varepsilon_i$'s $(4+\eta_1)$-th order moments is given in (C4)-(i) for  some  finite constant $\eta_1>0$. In addition, assume that $\rmE[\varepsilon_i^2|\mF_{\bmX,\bmW}]=\sigma_0^2$ for all $i=1,\cdots,n$,  $\bmS(\lambda)=\bmI_n-\lambda \bmW$ is nonsingular uniformly over $\lambda$ in a compact parameter space $\Lambda$ almost surely, and the true parameter value $\lambda_0$ is in the interior of $\Lambda$.
\end{as}

  The assumptions for the true parameter value $\lambda_0$ and $\bmS(\lambda)$ are the same as those in the classical spatial econometrics literature (see, e.g., \citealp{lee2004asymptotic} and \citealp{lesage2009introduction}),  but the assumption for $\bmS(\lambda)$ needs to hold almost surely for random $\bmS(\lambda)$. In addition, we impose the same conditions of Theorems \ref{tm:LLN} and  \ref{tm:CLT}  for the random errors $\varepsilon_i$ such that we can apply those theorems in developing the asymptotic normality of $\hat\bmtheta$ later.  We further assume that the random errors 
 have the same conditional variance. This assumption can lead to the  homoskedasticity of the classical SAR model (\ref{eq:SAR0}) 
 by the tower property of conditional expectation.

Before we introduce the next assumption, we introduce some notation below.  For a generic matrix $\bmG=( g_{k_1k_2})_{m\times q}$, if
$g_{k_1k_2}$ is a function of $\lambda\in \Lambda$, namely $\bmG=\bmG(\lambda)=\big( g_{k_1k_2}(\lambda)\big)_{m\times q}$, then we denote $\sup_{\lambda\in \Lambda}|\bmG(\lambda)|=\big( \sup_{\lambda\in \Lambda}|g_{k_1k_2}(\lambda)|\big)_{m\times q}$.  Based on this notation, we introduce  the assumption  associated to the random  weights matrix $\bmW$.

\begin{as}\label{as:2} Assume that   there exists $\eta_2>8/\eta_1$ such that
\ben
\left.\beal
&\max\biggl\{\sup_{n\geq 1}\lrnormmm {\bmW}_{L^{\delta_0(2+\eta_2)},1}, \sup_{n\geq 1}\lrnormmm {\bmW}_{L^{\delta_0(2+\eta_2)},\infty},\\
&\sup_{n\geq 1}\lrnormmm {\sup_{\lambda\in\Lambda}\lak \bmS^{-1}(\lambda)\rak}_{L^{\delta_1(2+\eta_2)},1}, \sup_{n\geq 1}\lrnormmm {\sup_{\lambda\in\Lambda}\lak \bmS^{-1}(\lambda)\rak}_{L^{\delta_1(2+\eta_2)},\infty}
\biggl\}<\infty,
\eeal\right.\label{eq:normmmc}
\een
for some  $\delta_0,\delta_1\in(0,\infty)$ that satisfy $
{4}{\delta_0^{-1}}+{2}{\delta_1^{-1}}\leq 1.
$
\end{as}

 If $\bmW$ is nonstochastic, Assumption \ref{as:2} reduces to the  uniform boundedness of  $\|\bmW\|_1$, $\|\bmW\|_\infty$, $\|\bmS^{-1}(\lambda)\|_1$ and $\|\bmS^{-1}(\lambda)\|_\infty$,  which is often considered in spatial econometrics (e.g., see Assumption 5 of \citealp{lee2004asymptotic}).  More interpretations of this assumption, including the relationships between $\normmm {\bmW}_{L^{\delta_0(2+\eta_2)},1}$  (or $\normmm {\bmW}_{L^{\delta_0(2+\eta_2)},\infty}$)  and $\|\bmW\|_1$ (or $\|\bmW\|_\infty$), are provided in Section S.5.1 of the supplementary material.  Based on these interpretations, the examination of Assumption \ref{as:2} is discussed in Section S.5.2 of the supplementary material. 
Since $\bmW$ is random, some moment conditions for random matrices $\bmW$ and $\bmS^{-1}(\lambda)$ via the definition of $\normmm{\cdot}$ in (\ref{eq:normmm}), are required in order to establish the asymptotic normality of $\hat\bmtheta$. These moment conditions also need to  be coordinated with $\varepsilon_i$'s $(4+\eta_1)$-th order moments in a way of those demonstrated in  Assumption \ref{as:2}. Similarly,  we next introduce the moment condition for the random regressors $\bmX$ below.

\begin{as}\label{as:3} For covariate matrix $\bmX=(\bmx_1,\cdots,\bmx_n)^\top=(x_{ij})_{n\times p}$, assume that there exists $\eta_3\in\big(4/(2+\eta_1),\eta_2\big)$ such that $\sup_{i,j}\rmE|x_{ij}|^{(2+\eta_2)(2+\eta_3)/(\eta_2-\eta_3)}<\infty$.\end{as}

 Assumption \ref{as:3} does not impose any independence conditions on $\bmx_i$ for $i=1,\cdots, n$, or $x_{ij}$ for $j=1,\cdots, p$. Instead, we only assume some moment condition on $x_{ij}$. This allows the dependence of $x_{ij}$ in both $i$ and $j$, and is more general than the nonstochastic $\bmX$ setting, since any nonstochastic constants $x_{ij}$ are mutually independent for $i=1,\cdots,n $ and $j=1,\cdots,p$. In addition,  this assumption allows the dependence between $\bmX$ and $\bmW$. 
 Moreover, the classical spatial econometrics literature (see, e.g., \citealp{kelejian2001asymptotic} and \citealp{lee2004asymptotic}) requires   the  condition of uniform boundedness in absolute value for the elements of $\bmX$  under the  nonstochastic $\bmX$ setting, i.e.,  $\max_{1\leq i\leq n;1\leq j\leq p}|x_{ij}|\leq C_x$ for some finite positive constant $C_x$. However,  if $x_{ij}$ are observations or realizations of random variables considered in some of the models and numerical studies of the above literature,  this condition may not be valid.   For example, if $x_{ij}\stackrel{iid}\sim N(0,\sigma_x^2)$, we have $\rmE(\max_{1\leq i\leq n;1\leq j\leq p}|x_{ij}|)/\sqrt{2\sigma_x^2\log(np)}\to 1$ as $n\to\infty$ (\citealp[Exercise 2.11, p. 53]{wainwright2019high}). Therefore, there exists a realization of the random variables $x_{ij}$ such that $\max_{1\leq i\leq n;1\leq j\leq p}|x_{ij}|/\sqrt{2\sigma_x^2\log(np)}\to 1$ (see, e.g., the similar technique in \citealp[p. 3]{vershynin2018high}), which indicates that this realization of $\max_{1\leq i\leq n;1\leq j\leq p}|x_{ij}|$ is not  uniformly bounded as $n\to\infty$. On the other hand, under such a random setting of $x_{ij}$, the moment condition in Assumption \ref{as:3} still holds, and hence Assumption \ref{as:3} releases the strong uniform boundedness condition for the elements of $\bmX$ in the literature.


Under Assumptions \ref{as:1} -- \ref{as:3} and using Theorem \ref{tm:LLN} (i), 
we subsequently obtain that
\[
\frac{1}{n}\rmE\lmk -\frac{\partial D_n(\bmtheta_0)}{\partial \bmtheta}\biggl{|}\mF_{\bmX,\bmW}\rmk=\bmo_{p+2},
\frac{1}{n}\Cov\lmk -\frac{\partial D_n(\bmtheta_0)}{\partial \bmtheta}\biggl{|}\mF_{\bmX,\bmW}\rmk=\bmV_{\rmS,n}+\bmOmega_{\rmS,n}, \textrm{ and }
\]
\[
\frac{1}{n}\rmE\lmk \frac{\partial^2  D_n(\bmtheta_0)}{\partial \bmtheta\partial\bmtheta^\top}\biggl{|}\mF_{\bmX,\bmW}\rmk=\bmU_{\rmS,n},
\]
 where the closed forms of $(p+2)\times (p+2)$ matrices $\bmV_{\rmS,n}$, $\bmOmega_{\rmS,n}$ and $\bmU_{\rmS,n}$ are  given by  (S.5), (S.6) and (S.7), respectively, in Section S.2 of the supplementary material. 
Based on  $\bmV_{\rmS, n}$, $\bmOmega_{\rmS,n}$ and $\bmU_{\rmS,n}$, we introduce the next assumption.

\begin{as}\label{as:4}  As $n\to\infty$, assume that $\bmV_{\rmS, n}\stackrel{\rmP}\longrightarrow \bmV_{\rmS}$, $\bmOmega_{\rmS,n}\stackrel{\rmP}\longrightarrow \bmOmega_{\rmS}$ and $\bmU_{\rmS,n}\stackrel{\rmP}\longrightarrow\bmU_{\rmS}$, where $\bmV_{\rmS}$, $\bmOmega_{\rmS}$ and $\bmU_{\rmS}$ are finite $(p+2)\times (p+2)$ matrices. In addition, assume that $\bmV_{\rmS}+\bmOmega_{\rmS}$ and $\bmU_{\rmS}$ are positive definite.
\end{as}

 Assumption \ref{as:4} is similar to Assumptions 6 and 8 of \citet{lee2004asymptotic} but under the random $\bmW$ and $\bmX$ setting, where  in this case $\bmV_{\rmS, n}$, $\bmOmega_{\rmS,n}$ and $\bmU_{\rmS,n}$ obtained in Section S.2 are all random  but not nonstochastic.  In addition, this assumption, together with Assumptions \ref{as:1} -- \ref{as:3},  leads to the asymptotic covariance matrix of  $-n^{-1/2}\partial D_n(\bmtheta_0)/\partial \bmtheta$, and the limit of $n^{-1}\partial^2 D_n(\bmtheta_0)/(\partial \bmtheta\partial \bmtheta^\top)$ in probability, both of which  are given in Lemma 11 of the supplementary material. Finally, we introduce the assumption for identification of $\bmtheta_0$. 

\begin{as} \label{as:5} As $n\to\infty$, assume that the limit of the $(p+1)\times(p+1)$ matrix  $n^{-1}\big( \bmX, \bmG_0\bmX\bmbeta_0\big)^\top \bmS(\lambda)\bmS(\lambda)^\top\big( \bmX, \bmG_0\bmX\bmbeta_0\big)$ in probability exists and is  nonsingular for  any $\lambda\in\Lambda$  and $\lambda\neq \lambda_0$, where  $\bmG_0=\bmW\bmS^{-1}_0$ and $\bmS_0=\bmS(\lambda_0)$.
\end{as}

 Assumption \ref{as:5}, together with Assumptions \ref{as:1} -- \ref{as:3}, lead to the identification of $\bmtheta_0$ and the consistency of  the quasi-score matching estimator $\hat\bmtheta$ given in Lemma 14 of the supplementary material. Based on Lemma 14, we demonstrate that when the QMLE  identifies $\bmtheta_0$ or is consistent, the quasi-score matching does so as well; see  Remark 2 in Section S.7 of the supplementary material.
The intuitive reasons for the consistency of $\hat\bmtheta$ are provided  in Remarks 1 and 3 in Section S.7 of the supplementary material.   Under heteroskedasticity, i.e., when $\rmE[\varepsilon_i^2|\mF_{\bmX,\bmW}]=\sigma_0^2$ in Assumption \ref{as:1} is violated, $\hat\bmtheta$ is inconsistent, whereas the 2SLS estimator \citep{kelejian1998generalized} remains consistent and robust;  however, the 2SLS estimation does not work in the case that $\rmE[\bmW \bmy|\mF_{\bmX,\bmW}]$ and $\bmX$ are linearly dependent. On the other hand, we show that the quasi-score matching estimation works in this case; see Section S.3 of the supplementary material.

Lemma 14, in conjunction with Assumption \ref{as:4}, yields
 the asymptotic normality of $\hat\bmtheta$ given in the theorem below.

\begin{tm}\label{tm:SM} Under  Assumptions \ref{as:1} -- \ref{as:5}, we have that, as $n\to\infty$,
\[
\sqrt{n}(\hat\bmtheta-\bmtheta_0)\stackrel{d}\longrightarrow N(\bmo_{p+2},\bmU_{\rmS}^{-1}\bmV_{\rmS}\bmU_{\rmS}^{-1}+\bmU_{\rmS}^{-1}\bmOmega_{\rmS}\bmU_{\rmS}^{-1}).
\]
\end{tm}

Theorem \ref{tm:SM} can  lead to the asymptotic variance of the quasi-score matching estimator $\hat\bmtheta$, and it is $n^{-1}(\bmU_{\rmS}^{-1}\bmV_{\rmS}\bmU_{\rmS}^{-1}+\bmU_{\rmS}^{-1}\bmOmega_{\rmS}\bmU_{\rmS}^{-1})$.  Although the quasi-score matching approach can alleviate the computational complexity of the QMLE when $n$ is large, we utilize the asymptotic variance obtained from Theorem \ref{tm:SM} to show that it trades off some estimation efficiency compared to the QMLE; see Corollary 1 in Section S.4.1 of the supplementary material. 
Nevertheless, based on the numerical estimation efficiencies demonstrated in various simulations in Section \ref{sec:simu}, we find that the quasi-score matching achieves efficiency close to the QMLE, especially for large $n$.  In the following subsection,  we  propose an approach to improve the asymptotic efficiency of the quasi-score matching estimators $\hat\bmbeta$ and $\hat\sigma^2$ while retaining the computational complexity as $n$ is large.

\subsection{Improving Asymptotic Efficiency for Estimation of $\bmbeta$ and $\sigma^2$}
\label{subsec:improve}

Recall that the concentrated likelihood approach for the QMLE leads to two estimating equations (\ref{eq:beta}) and (\ref{eq:sigma2}) that are used to estimate $\bmbeta$ and $\sigma^2$ given $\lambda$. As a consequence, if we substitute $\lambda$ in  (\ref{eq:beta}) and (\ref{eq:sigma2}) with the  quasi-score matching estimator $\hat\lambda$, we can have  two new estimators of $\bmbeta$ and $\sigma^2$:
\ben
\tilde\bmbeta(\hat\lambda)=(\bmX^\top \bmX)^{-1}\bmX^\top \bmS(\hat\lambda)\bmy,\textrm{ and }
\een
\ben
\tilde\sigma^2(\hat\lambda)=\frac{1}{n} \bmy^\top\bmS(\hat\lambda)^\top \bmM_{\bmX} \bmS(\hat\lambda)\bmy.
\een
One can see that the computational complexity of obtaining these estimators is not increased compared to the calculations of the quasi-score matching estimators $\hat\bmbeta$ and $\hat\sigma^2$ in (\ref{eq:sm}). However, since the information of the QMLE's estimating equations is involved in these new estimators, one can expect an improvement of the efficiency  for the estimation of $\bmbeta$ and $\sigma^2$.

In what follows,  we study the asymptotic properties of the efficiency improved estimators $\hat{\tilde\bmbeta}=\tilde\bmbeta(\hat\lambda)$ and $\hat{\tilde\sigma}^2=\tilde\sigma^2(\hat\lambda)$ under the random $\bmW$ and $\bmX$ setting. In particular, we show a more general result for  the estimator  $\hat{\tilde\bmtheta}=(\hat\lambda, \hat{\tilde\bmbeta}^\top,\hat{\tilde\sigma}^2)^\top$, where $( \hat{\tilde\bmbeta}^\top,\hat{\tilde\sigma}^2)^\top$ is just a sub-vector of  $\hat{\tilde\bmtheta}$. Since the efficiency improved estimator  $\hat{\tilde\bmtheta}=(\hat\lambda, \hat{\tilde\bmbeta}^\top,\hat{\tilde\sigma}^2)^\top$ includes the information of both the quasi-score matching estimator $\hat\lambda$ and the QMLE's estimating equations  (\ref{eq:beta}) and (\ref{eq:sigma2}), the asymptotic distribution of $\hat{\tilde\bmtheta}$ involves the moments of both ${\partial D_n(\bmtheta_0)}/{\partial \bmtheta}$ and ${\partial \ell_n(\bmtheta_0)}/{\partial \bmtheta}$, where $D_n(\bmtheta)$ is in (\ref{eq:obj})  from the quasi-score matching, and $\ell_n(\bmtheta)$ is in (\ref{eq:likelihood}) from the QMLE. Specifically,  we use Theorem \ref{tm:LLN}  (i)  and Assumptions \ref{as:1} -- \ref{as:3} to obtain  that
\[
\frac{1}{n}\Cov\lmk \begin{pmatrix}-\frac{\partial D_n(\bmtheta_0)}{\partial \bmtheta}\\\frac{\partial \ell_n(\bmtheta_0)}{\partial \bmtheta}\end{pmatrix}\Biggl{|}\mF_{\bmX,\bmW}\rmk=\bmV_{n}+\bmOmega_{n},
\]
 where the closed forms of $2(p+2)\times 2(p+2)$ matrices  $\bmV_{n}$ and $\bmOmega_{n}$ are  given by (S.8) in Section S.2 of the supplementary material. Similar to Theorem \ref{tm:SM}, to show the asymptotic normality of $\hat{\tilde\bmtheta}$, we require an extra assumption on  $\bmV_n$ and $\bmOmega_n$, which is given below.

\begin{as:4'}  As $n\to\infty$, assume that \[\bmV_{n}\stackrel{\rmP}\longrightarrow \bmV=\begin{pmatrix}\bmV_{\rmS}&\bmV_{\rmS\rmM}\\\bmV_{\rmS\rmM}^\top &\bmV_{\rmM}\end{pmatrix}, \textrm{ and }\bmOmega_{n}\stackrel{\rmP}\longrightarrow \bmOmega=\begin{pmatrix}\bmOmega_{\rmS}&\bmOmega_{\rmS\rmM}\\\bmOmega_{\rmS\rmM}^\top &\bmOmega_{\rmM}\end{pmatrix},\]where $\bmV$ and $\bmOmega$ are finite $2(p+2)\times 2(p+2)$ matrices such that $\bmV+\bmOmega$ and $\bmV$ are  positive definite, $\bmV_{\rmS}$ and $\bmOmega_{\rmS}$  are $(p+2)\times (p+2)$ matrices  defined in Assumption \ref{as:4}, and $\bmV_{\rmM}$, $\bmOmega_{\rmM}$, $\bmV_{\rmS\rmM}$ and $\bmOmega_{\rmS\rmM}$  are finite $(p+2)\times (p+2)$ matrices.
\end{as:4'}

\noindent{The examination of Assumption \ref{as:4}$'$ and Assumption \ref{as:4} above is discussed in Section S.6 of the supplementary material.} 

Based on this assumption and $\bmU_{\rmS}$ in  Assumption \ref{as:4}, let
\be\label{eq:VM}
\bmV_{\rmM}=\begin{pmatrix} V_{\rmM,\lambda,\lambda} &\bmV_{\rmM,\lambda,-\lambda}\\
\bmV_{\rmM,-\lambda,\lambda}&\bmV_{\rmM,-\lambda,-\lambda}\\
\end{pmatrix},
\ee
for $V_{\rmM,\lambda,\lambda}\in\mathbb{R}$, $\bmV_{\rmM,\lambda,-\lambda}\in\mathbb{R}^{1\times (p+1)}$, $\bmV_{\rmM,-\lambda,\lambda}\in\mathbb{R}^{(p+1)\times 1}$ and $\bmV_{\rmM,-\lambda,-\lambda}\in\mathbb{R}^{(p+1)\times (p+1)}$, and denote
\be\label{eq:Xi}
\bmXi=\begin{pmatrix} \lsk 1,\bmo_{1\times (p+1)}\rsk\bmU_{\rmS}^{-1}&\bmo_{1\times (p+2)}\\
- \bmV_{\rmM,-\lambda,-\lambda}^{-1}\bmV_{\rmM,-\lambda,\lambda} \lsk 1,\bmo_{1\times (p+1)}\rsk\bmU_{\rmS}^{-1} & \bmV_{\rmM,-\lambda,-\lambda}^{-1}(\bmo_{p+1},\bmI_{p+1})
\end{pmatrix}\in\mathbb{R}^{(p+2)\times 2(p+2)}
\ee
The notation, together with  Assumptions \ref{as:1} -- \ref{as:5} and \ref{as:4}$'$,  leads to the asymptotic normality of $\hat{\tilde\bmtheta}$ given below.

\begin{tm}\label{tm:I} Under  Assumptions \ref{as:1} -- \ref{as:5} and \ref{as:4}$\,'$, we have that, as $n\to\infty$,
\[
\sqrt{n}\lsk \hat{\tilde\bmtheta}-\bmtheta_0\rsk\stackrel{d}\longrightarrow N\lsk \bmo_{p+2},\bmXi\bmV\bmXi^\top+\bmXi\bmOmega\bmXi^\top\rsk.
\]
\end{tm}

 Based on this theorem, Theorem \ref{tm:SM}, and Lemma 15 in Section S.7 of the supplementary material, we show that the QMLE $\tilde\bmtheta$ is asymptotically more efficient than either the quasi-score matching estimator $\hat\bmtheta=(\hat\lambda,\hat\bmbeta^\top,\hat\sigma^2)^\top$ or the efficiency improved estimator $\hat{\tilde\bmtheta}=(\hat\lambda, \hat{\tilde\bmbeta}^\top,\hat{\tilde\sigma}^2)^\top$; see Section S.4.1 and Corollary 1  of the supplementary material. In addition, Section S.4.1 provides numerical justification that  the asymptotic efficiency of $\hat{\tilde\bmtheta}$ is better than that of   $\hat\bmtheta$, and our simulation studies in Section \ref{sec:simu} further confirm this result.  As suggested by an anonymous reviewer, we also prove that the efficiency improved estimator   is asymptotically more efficient than  the best 2SLS estimator \citep{lee2003best}  under some conditions; see Section S.4.2 and Corollary 2  of the supplementary material.

 Note that both the calculations of $\hat{\tilde\bmtheta}$ and $\hat\bmtheta$ are  simpler and faster than that of $\tilde\bmtheta$ as $n$ is large, while the asymptotic variances of the three estimators are all of order $O(n^{-1})$.  We next provide the simulation studies to 
 compare the performance of $\hat\bmtheta$, $\hat{\tilde\bmtheta}$ and $\tilde\bmtheta$ in Section \ref{sec:simu} below.

\section{Simulation}
\label{sec:simu}

In this section, we conduct various simulation studies to  evaluate the finite sample performance of the quasi-score matching estimator $\hat\bmtheta=(\hat\lambda,\hat\bmbeta^\top,\hat\sigma^2)^\top$ and the efficiency improved estimator $\hat{\tilde\bmtheta}=(\hat\lambda, \hat{\tilde\bmbeta}^\top,\hat{\tilde\sigma}^2)^\top$. Subsequently, we compare them to the QMLE $\tilde\bmtheta=(\tilde\lambda,\tilde\bmbeta^\top,\tilde\sigma^2)^\top$. It is worth noting that the efficiency improved estimator $\hat{\tilde\bmtheta}$ is a variant of the quasi-score matching estimator, whose calculation relies on obtaining the estimate $\hat\lambda$ in $\hat\bmtheta$. Without causing confusion, we call both $\hat\bmtheta$ and $\hat{\tilde\bmtheta}$ quasi-score matching estimators in the following studies.  All the tables in this section, except Table  \ref{tb:1},  are provided in Section S.9 of the supplementary material to save space.

In particular, the simulated data are generated from model (\ref{eq:SAR0}),  where  $\lambda_0=0.3$, $\bmbeta_0=(2,1)^\top$,  $\bmX=(x_{ij})_{n\times 2}$ with $x_{i1}\equiv 1$ and $x_{i2}$ being independently generated from the standard normal distribution for $i=1,\cdots, n$,  and the random errors $\varepsilon_i$ are independently generated from the standard normal distribution and the mixture normal distribution $0.9N(0,5/9)+0.1N(0, 5)$, respectively. Hence, $\sigma_0^2=1$. Moreover,  we follow the recent literature of network data analysis (see, e.g., \citealp{zhu2020multivariate}) to construct the spatial weights matrix $\bmW=(w_{ij})_{n\times n}$  by the row-normalization of  network adjacency matrix $\bmA=(a_{ij})_{n\times n}$, i.e., $w_{ij}=a_{ij}/\sum_{j'=1}^na_{ij'}$,  where $a_{ij}$ is either one or zero for $i,j=1,\cdots, n$, and the diagonals $a_{ii}$ are all zeros. In this study, we consider two types of $\bmA$ which are generated from
\begin{itemize}
\item Simple network (Bernoulli-type): The off-diagonal elements $a_{ij}$  are independently generated from the Bernoulli distribution with probability $5/n$; and
\item Stochastic block model  (SBM-type): We first randomly assign a block label varying from 1 to 3  with the equal probability to indices $i=1,\cdots, n$. Next we generate off-diagonal elements $a_{ij}$ with probability $\rmP(a_{ij}=1)=10/n$ if indices $i$ and $j$ have the same block label, and probability $\rmP(a_{ij}=1)=5/n^{1.1}$ otherwise.
\end{itemize}

\noindent For the convenience of referring to these two settings of simulation, we name the weights matrices constructed by the simple network $\bmA$ and the SBM $\bmA$, respectively, as the Bernoulli-type and SBM-type weights matrices.  It is worth noting that both the  Bernoulli-type and SBM-type weights matrices are asymmetric under the above settings. Lastly, we consider four sample sizes  $n=500$, 1,000, 5,000, and 10,000 in the simulation studies.

 For each of the above simulation settings, we generate $r=1,\cdots,1000$ replications of simulated data.
 In the $r$-th replication,  we estimate the parameter  $ \bmtheta_0=(\lambda_0,\bmbeta_0^\top, \sigma_0^2)^\top=(\theta_{1,0},\cdots,\theta_{4,0})^\top$ by $\hat\bmtheta=(\hat\lambda,\hat\bmbeta^\top, \hat\sigma^2)^\top$,  $\hat{\tilde\bmtheta}=(\hat\lambda,\hat{\tilde\bmbeta}^\top, \hat{\tilde\sigma}^2)^\top$, and $\tilde\bmtheta=(\tilde\lambda,\tilde\bmbeta^\top, \tilde\sigma^2)^\top$. For convenience, we denote the estimates in the $r$-th replication by $\hat\bmtheta^{(r)}=(\hat\theta_1^{(r)},\cdots,\hat\theta_4^{(r)})^\top$,  $\hat{\tilde\bmtheta}^{(r)}=(\hat{\tilde \theta}_1^{(r)},\cdots,\hat{\tilde \theta}_4^{(r)})^\top$, and $\tilde\bmtheta^{(r)}=(\tilde\theta_1^{(r)},\cdots,\tilde\theta_4^{(r)})^\top$, respectively. To evaluate the  performance of these estimates,  we consider three measurements:  the empirical bias (BIAS) $1000^{-1}\sum_{r=1}^{1000}(\hat\theta^{(r)}_j$ $- \theta_{j,0})$, the empirical standard deviation (SD) $\{1000^{-1}\sum_{r=1}^{1000}(\hat\theta_j^{(r)}- 1000^{-1}\sum_{r=1}^{1000}\hat\theta_j^{(r)})^2\}^{1/2}$, and the root mean  squared error (RMSE) $(\textrm{BIAS}^2$ $+\textrm{SD}^2)^{1/2}$, for $\hat\bmtheta$ and $j=1,\cdots, 4$; and similarly for $\hat{\tilde\bmtheta}$ and  $\tilde\bmtheta$. To assess the computational complexities of obtaining these estimates,  we calculate the average computation times $t_{\rmS}=1000^{-1}\sum_{r=1}^{1000}t_{\rmS}^{(r)}$ and $t_{\rmM}=1000^{-1}\sum_{r=1}^{1000}t_{\rmM}^{(r)}$, where 
 for the $r$-th replication,  $t_{\rmS}^{(r)}$ is the computation time of obtaining the quasi-score matching  estimates $\hat\bmtheta$ and $\hat{\tilde\bmtheta}$, and $t_{\rmM}^{(r)}$ is the computation time of obtaining the QMLE $\tilde\bmtheta$.

Under   the standard normal random errors, Tables \ref{tb:1} and S.1 present the simulation results for the settings of Bernoulli-type weights matrix and SBM-type weights matrix, respectively, and reveal four important findings. (I) In general, the magnitudes of BIAS and SD become smaller for the two quasi-score matching estimates and the QMLE as $n$ gets larger.  It is not surprising that the RMSE shows the same pattern, which indicates the consistency of all three estimators. (II) The difference of SD and RMSE between the quasi-score matching estimate $\hat\bmtheta$ and the QMLE $\tilde\bmtheta$ is small, and gets even smaller as $n$ increases. In Table \ref{tb:1}, for example,  the RMSE difference between 
$\hat\lambda$ and  $\tilde\lambda$ is $0.61\times 10^{-2}$ for $n=500$, and is $0.11\times 10^{-2}$ for $n=10,000$. Similarly, the RMSE difference between  $\hat\beta_2$ and $\tilde\beta_2$ reduces from $0.14\times 10^{-2}$ to $0.03\times 10^{-2}$ as $n$ increases from 500 to 10,000. (III)  The  SD and RMSE of the efficiency improved estimates $\hat {\tilde\beta}_1$, $\hat {\tilde\beta}_2$ and $\hat {\tilde\sigma}^2$ are indeed smaller than those of the quasi-score matching estimates $\hat {\beta}_1$, $\hat {\beta}_2$ and $\hat {\sigma}^2$,  which confirms the remarks after Theorem \ref{tm:I}. In addition,  the difference of SD and RMSE between the efficiency improved estimate $\hat {\tilde\beta}_2$ and the QMLE $\tilde \beta_2$  is not visible  in  Tables \ref{tb:1} and S.1's decimal places ($0.01\times 10^{-2}$) under the smallest sample size $n=500$, not to mention under other larger settings of $n$. Similar findings can be found for  $\hat {\tilde\sigma}^2$ and  $\tilde \sigma^2$.  (IV) In terms of the computation time, calculating the quasi-score matching estimates $\hat\bmtheta$ and $\hat{\tilde\bmtheta}$ takes much less than calculating the QMLE $\tilde\bmtheta$. For instance, Table \ref{tb:1} shows that it takes 1051.27 seconds on average to obtain the QMLE for $n=$ 10,000, while the computation of  the quasi-score matching estimates only requires 0.04 seconds (1/26119  of 1051.27). In Table S.1, the average computation time  becomes  985.61 seconds for the QMLE, but is only  0.059 seconds (1/16691 of  985.61) for the quasi-score matching estimates. In summary, Tables \ref{tb:1}  and  S.1 indicate the huge computation gain of the quasi-score matching estimation for a large $n$, while its efficiency sacrifice is minor, and is negligible for non-intercept regression coefficient $\beta_2$ and variance of random errors $\sigma^2$.

\begin{table}[htbp!]
\renewcommand{\arraystretch}{0.75}
\setlength{\tabcolsep}{4pt}
\caption{Simulation results of the QMLE and the quasi-score matching estimates under $\lambda_0=0.3$, the standard normal random errors and the Bernoulli-type weights matrix.} 
\label{tb:1}
\vspace{-20pt}
\begin{center}
\scalebox{0.8}{
\begin{tabular}{|c|c|rrrr|rrrrrrr|}
  \hline
  & & \multicolumn{4}{c|}{QMLE}    &     \multicolumn{7}{c|}{Quasi-Score Matching} \\
\cline{3-6}\cline{7-13}
 $n$  &\multicolumn{1}{c|}{}& \multicolumn{1}{c}{$\tilde \lambda$} & \multicolumn{1}{c}{{$\tilde \beta_1$}} & \multicolumn{1}{c}{{$\tilde\beta_2$}} & \multicolumn{1}{c}{$\tilde\sigma^2$} & \multicolumn{1}{|c}{$\hat \lambda$} & \multicolumn{1}{c}{$\hat\beta_1$} & \multicolumn{1}{c}{$\hat\beta_2$} & \multicolumn{1}{c}{$\hat\sigma^2$}  & \multicolumn{1}{c}{$ \hat{\tilde{\beta_1}}$} & \multicolumn{1}{c}{$\hat{\tilde{\beta_2}}$} & \multicolumn{1}{c|}{$\hat{\tilde{\sigma}}^2$} \\
   \hline
   500 &BIAS  & -0.80 & 2.35 & 0.13 & -0.66 & -0.38 & 1.22 & 0.22 & -0.62 & 1.17 & 0.14 & -0.62 \\ 
   & SD & 6.03 & 17.67 & 4.47 & 6.30 & 6.68 & 19.55 & 4.60 & 6.37 & 19.48 & 4.47 & 6.30 \\ 
   &RMSE & 6.08 & 17.82 & 4.47 & 6.34 & 6.69 & 19.59 & 4.61 & 6.40 & 19.51 & 4.47 & 6.33 \\ 
   \cline{2-13}
  & time (seconds)& \multicolumn{4}{c|}{ $t_{\text{M}}=0.1192$} &  \multicolumn{7}{c|}{$t_{\text{S}}=0.0052  \quad ( t_{\text{M}}/t_{\text{S}}=22.88)$ }\\
   \hline
  1,000 &BIAS& -0.28 & 0.80 & 0.08 & -0.19 & -0.01 & 0.05 & 0.10 & -0.17 & 0.05 & 0.08 & -0.17 \\ 
   &SD& 4.05 & 11.93 & 3.11 & 4.50 & 4.26 & 12.56 & 3.22 & 4.54 & 12.54 & 3.11 & 4.50 \\ 
   &RMSE& 4.06 & 11.96 & 3.11 & 4.51 & 4.26 & 12.56 & 3.22 & 4.55 & 12.54 & 3.11 & 4.51 \\ 
    \cline{2-13}
   & time (seconds) & \multicolumn{4}{c|}{ $t_{\text{M}}=0.6864$} & \multicolumn{7}{c|}{$t_{\text{S}}=0.0061  \quad (t_{\text{M}}/t_{\text{S}}=111.97)$}\\
   \hline
  5,000 &BIAS& -0.03 & 0.06 & -0.00 & -0.11 & -0.00 & -0.02 & -0.02 & -0.10 & -0.03 & -0.00 & -0.11 \\ 
   &SD& 1.84 & 5.44 & 1.44 & 1.99 & 1.97 & 5.77 & 1.47 & 1.99 & 5.78 & 1.44 & 1.99 \\ 
   &RMSE& 1.84 & 5.44 & 1.44 & 1.99 & 1.97 & 5.77 & 1.47 & 2.00 & 5.78 & 1.44 & 1.99 \\
    \cline{2-13}
   &  time (seconds) &  \multicolumn{4}{c|}{ $t_{\text{M}}=108.9210 $} & \multicolumn{7}{c|}{$t_{\text{S}}=0.0187 \quad (t_{\text{M}}/t_{\text{S}}=5815.32)$}\\ 
   \hline
  10,000& BIAS& -0.05 & 0.18 & -0.00 & -0.04 & -0.01 & 0.09 & 0.01 & -0.05 & 0.07 & -0.00 & -0.04 \\ 
   &SD& 1.34 & 3.86 & 0.98 & 1.38 & 1.45 & 4.18 & 1.01 & 1.39 & 4.17 & 0.98 & 1.38 \\ 
  &RMSE& 1.34 & 3.86 & 0.98 & 1.38 & 1.45 & 4.18 & 1.01 & 1.39 & 4.17 & 0.98 & 1.38 \\
    \cline{2-13} 
  & time (seconds) &  \multicolumn{4}{c|}{ $t_{\text{M}}=1051.2729$} & \multicolumn{7}{c|}{$t_{\text{S}}=0.0402 \quad (t_{\text{M}}/t_{\text{S}}=26118.58)$ }\\
  \hline
 \multicolumn{13}{l}{Note:  The BIAS, SD and RMSE in this table  
  are 100 times their actual values.} 
\end{tabular}}
\end{center}
\vspace{-14pt}
\end{table}

 After analyzing the estimation performance under $\lambda_0=0.3$ and  the standard normal random errors, we examine the performance of the parameter estimates and computation times under other settings. To save space, these simulation results are relegated to  Section S.9.1 of 
the supplementary material.  Specifically, Tables S.2 -- S.3 in the supplementary material report the results when the random errors are generated from the mixture normal distribution, but $\lambda_0$ remains to be 0.3. In addition, Tables S.4 -- S.5 and Tables S.6 -- S.7 in the supplementary material present the results under settings of $\lambda_0=0.5$ and 0.7, respectively, and the rest of the simulation settings 
are the same as those in Tables \ref{tb:1} and S.2.   
These tables show qualitatively similar findings to those in Tables \ref{tb:1}  and  S.1.

 In addition to these results, additional simulation studies according to anonymous referees’ suggestions are presented in Sections S.9.2  -- S.9.5 of the supplementary material. Specifically, Section S.9.2 generates the QQ plots for our proposed estimates in simulations and confirms  the asymptotic normality  in Theorems \ref{tm:SM} -- \ref{tm:I}.  
 Section S.9.3.1 examines the convergence rate $n^{-1}$ in the asymptotic variances of estimators, and  demonstrates that the convergence rate of the RMSEs under the Bernoulli-type and SBM-type weights matrices above is approximately $1/\sqrt{n}$. This section also introduces an alternative SBM setting that generates a denser network adjacency matrix $\bmA$, whose resulting weights matrix violates Assumptions \ref{as:4} and \ref{as:4}$'$, leading to a slower RMSE convergence rate. Section S.9.3.2 shows that the SDs of the quasi-score matching estimates $\hat\bmtheta$ and the efficiency-improved estimates $\hat{\tilde\bmtheta}$, are close to their corresponding theoretical asymptotic standard deviations obtained from Theorems \ref{tm:SM} -- \ref{tm:I}.  Section S.9.4 demonstrates the computational advantage of the quasi-score matching estimation over the QMLE when the determinant in the QMLE log-likelihood function is  computed using the eigenvalue decomposition algorithm. Section S.9.5 compares our proposed quasi-score matching method with the GMM estimation by \citet{lee2007gmm}, 
 the LSE  by \citet{huang2019least}, the 2SLS estimation  by \citet{kelejian1998generalized}, 
 the best 2SLS estimation by \citet{lee2003best} and the one-step estimation by \citet{gupta2023efficient}, and shows that our method is computationally faster than all of these methods except 2SLS, particularly for large $n$, where 2SLS is only slightly faster but exhibits lower estimation efficiency.

\section{Case Study}
\label{sec:casestudy}

 In this study, we consider the social network experiment  conducted in \citet{paluck2016changing}. The purpose of the experiment was to test the theories that individuals attend to the behavior of certain people in their community, and to understand what is socially normative to adjust individuals' own behavior in their responses. This experiment was conducted over a school year from September 2012 to June 2013, and  28 middle schools were randomly selected to receive an anti-conflict intervention  (treatment) via organizing an education workshop about conflicts from 56 middle schools in New Jersey.  More details can be found in \citet{paluck2016changing}.

 To measure students' social connections,  a survey was conducted at the start of the school year. 
 In this survey, students were asked to nominate up to ten peers with whom they spent most time in the last few weeks. Based on  this survey, if student $i$ nominates student $j$, for $i,j=1,\cdots,n$, there is an edge from student $i$ to student $j$ in the social network of $n$ students.  Accordingly, $a_{ij}$ in the associated network adjacency matrix $\bmA=(a_{ij})_{n\times n}$ is set to be 1; otherwise, $a_{ij}$ is set to be 0. For completeness, $a_{ii}$ are set to be zeros for $i=1,\cdots,n$.  It is worth noting that the nominations are not symmetric, and hence the constructed adjacency matrix $\bmA$ is asymmetric.  These students' nominations in the  survey result in a realization of the row-normalized adjacency matrix $\bmW=(w_{ij})_{n\times n}$ as the weights matrix in  the SAR model (\ref{eq:SAR0}), where $w_{ij}=a_{ij}/\sum_{j'=1}^na_{ij'}$ for $i,j=1,\cdots,n$.  Alongside the edge information between any two students $i,j\in\{1,\cdots, n\}$, there were various questions in a  second survey after the experiment about each student $i$'s perception of their school's conflict situations, as well as their demographics. Aside from the  surveys, schools' administrative records collected each student $i$'s  conflict-related disciplinary events across the school year.

To utilize the SAR model (\ref{eq:SAR0})  to study the impacts of the anti-conflict intervention and students' social network, on their perception of the conflict situations at school, we construct the response variable $y_i$, $i=1,\cdots, n$, as follow. First, there were 13 questions in  the second survey for students' view on their school's conflict situations. Eight of them were positively worded questions. For example, ``How many students at this school think it's good to be friendly and nice with all students at this school, no matter who?'' The remaining five were negatively worded questions. For instance, ``How many students at this school think it's funny, or not a big deal, to post something mean about someone online?"  For each of these questions, students were asked to give a score from 0 (``almost nobody'') to 5 (``almost everyone''). Second,  we  reverse the scores of those negatively worded questions.  By doing so, a high score indicates a more friendly and positive school environment consistently for all the 13 questions. Lastly, the response variable $y_i$ is the average score of these 13 questions, which measures the overall feeling of student $i$ towards their school's conflict atmosphere. Preliminary results show that the distribution of $y_i$s is roughly Gaussian.

In addition to the response variable, we construct the regressors $\bmx_i=(1, x_{i2},\cdots, x_{ip})^\top$ for students $i=1,\cdots,n$ given below.  We first use a 0/1 binary variable to indicate the treatment, i.e., if student $i$ was from the 28 treated schools, then $x_{i2}=1$; otherwise, $x_{i2}=0$. The other regressors $(x_{i3},\cdots, x_{ip})^\top$ consist of the  demographic variables: gender (girl or boy), grade, race (White, Black, Hispanic/Latino, Asian-American, South Asian or Other), whether student $i$ lived with both parents, whether student $i$ was a return student from the previous year, and the number of times that student $i$ was disciplined for a conflict during the school year in which the intervention was conducted.

We  then fit the SAR model (\ref{eq:SAR0}) with the weights matrix $\bmW$ constructed  based on the  survey conducted at the start of the school year, while we remove the observations with missing values or no network connections in the model fitting, which results in $n=$ 14,732 students in total.  We follow Section \ref{sec:simu} to obtain the QMLE $\tilde\bmtheta=(\tilde\lambda,\tilde\bmbeta^\top, \tilde\sigma^2)^\top$, the quasi-score matching estimate $\hat\bmtheta=(\hat\lambda,\hat\bmbeta^\top, \hat\sigma^2)^\top$ and the efficiency improved estimate $\hat{\tilde\bmtheta}=(\hat\lambda,\hat{\tilde\bmbeta}^\top, \hat{\tilde\sigma}^2)^\top$.  The estimation results are summarized in Table \ref{tb:data1}.  Overall, the three estimates and their standard errors are similar. In particular, the efficiency improved estimate $\hat{\tilde \bmtheta}$ is  numerically closer to the QMLE $\tilde\bmtheta$ compared to the quasi-score matching estimate $\hat\bmtheta$. Moreover, the standard errors of $\hat{\tilde\bmbeta}$ and $ \hat{\tilde\sigma}^2$ are almost identical to those of $\tilde\bmbeta$  and $\tilde\sigma^2$, which is consistent with the findings in the simulation studies.

\begin{table}[htbp!]
\renewcommand{\arraystretch}{1.0}
\setlength{\tabcolsep}{4pt}
\caption{Estimates, standard errors and $p$-values of the SAR model with  $\bmW$ constructed based on the  survey conducted at the start of the school year.} 
\label{tb:data1}
\vspace{-20pt}
\begin{center}
\scalebox{0.8}{
\begin{tabular}{|c|rr|rr|rr|}
  \hline
  & \multicolumn{2}{c|}{QMLE $\tilde \bmtheta$}    &    \multicolumn{2}{c|}{Quasi-Score Matching $\hat \bmtheta$} &\multicolumn{2}{c|}{Quasi-Score Matching $\hat{\tilde\bmtheta}$} \\
\cline{2-3}\cline{4-5}\cline{6-7}
&coef. (std.error)&$p$-value&coef. (std.error)&$p$-value&coef. (std.error)&$p$-value\\
\hline
$\lambda$& 0.1993 (0.0121)&$<10^{-4}$&0.1913 (0.0127)&$<10^{-4}$&0.1913 (0.0127)&$<10^{-4}$\\
 \hline
Intercept &2.6842 (0.0500) &$<10^{-4}$&2.7175 (0.0517)&$<10^{-4}$&2.7104 (0.0515)&$<10^{-4}$\\
Treatment &-0.0748 (0.0135) &$<10^{-4}$&-0.0775 (0.0137)&$<10^{-4}$&-0.0755 (0.0136)&$<10^{-4}$\\
Gender:boy &0.0247 (0.0135) &0.0670&0.0321 (0.0137)&0.0189&0.0245 (0.0135)&0.0693\\
Grade &-0.1388 (0.0084) &$<10^{-4}$&-0.1383 (0.0085)&$<10^{-4}$&-0.1401 (0.0084)&$<10^{-4}$\\
Race:White &0.0527 (0.0212)&0.0131&0.0467 (0.0215)&0.0299&0.0532 (0.0212)&0.0122\\
Race:Black&-0.1417 (0.0279)&$<10^{-4}$&-0.1270 (0.0283)&$<10^{-4}$&-0.1429 (0.0279)&$<10^{-4}$\\
Race:Hispanic/Latino&-0.0309 (0.0228)&0.1753&-0.0321 (0.0231)&0.1650&-0.0312 (0.0228)&0.1724\\
Race:Asian-American &0.0693 (0.0343) &0.0433&0.0599 (0.0347)&0.0844&0.0699 (0.0343)&0.0418\\
Race:South Asian &0.0732 (0.0604) &0.2251&0.0499 (0.0609)&0.4126&0.0737 (0.0604)&0.2219\\
Live with both parents &0.0528 (0.0158) &0.0008&0.0437 (0.0159)&0.0061&0.0533 (0.0158)&0.0007\\
Return student&-0.0299 (0.0172) &0.0826&-0.0310 (0.0177)&0.0789&-0.0301 (0.0172)&0.0807\\
Discipline event counts & -0.0189 (0.0028)&$<10^{-4}$&-0.0175 (0.0029)&$<10^{-4}$&-0.0190 (0.0028)&$<10^{-4}$\\
 \hline
$\sigma^2$&0.6564 (0.0078) &$<10^{-4}$&0.6543 (0.0078)&$<10^{-4}$&0.6567 (0.0078)&$<10^{-4}$\\
  \hline
\end{tabular}}
\end{center}
\vspace{-30pt}
\end{table}

 In terms of testing the significance of parameters, the three estimation approaches lead to the same conclusions for parameters $\lambda$,  $\sigma^2$ and $\bmbeta$ at the significance level 0.10, but with two exceptions for $\bmbeta$ at the significance level 0.05:  the coefficients of gender and   Asian-American. In these two exceptions, the QMLE and the efficiency improved estimate  both imply the same conclusions, i.e.,  the coefficient of gender is not significant and the coefficient of  Asian-American is significant. On the other hand, the quasi-score matching estimate gives different conclusions based on the $p$-values. This finding is sensible since the efficiency improved estimator has a smaller estimation error for $\bmbeta$ compared to the quasi-score matching estimator, which is demonstrated in the simulation studies of Section \ref{sec:simu}. It is worth noting that the efficiency improved estimator is designed to alleviate the computation for a large $n$, though in this study we also calculate the QMLE as a benchmark. Based on the above comparisons, we recommend using the efficiency improved estimate in practice for fitting the SAR model with a large $n$.

We subsequently focus on interpreting the results of the efficiency improved estimate $\hat{\tilde \bmtheta}$ in Table \ref{tb:data1}, which show seven interesting findings at the significance level 0.05. First, the positive significance of the spatial autoregressive coefficient $\lambda$ indicates that the social network effects exist and the students' perception of their school's conflict situations highly depends on the view of their connected peers.  Second, the treatment is significantly negative. This is sensible because students became more aware of their school's conflicts than what they used to be due to the organized education workshop about conflicts. This result is consistent with the findings in \citet{paluck2016changing}, in which  they find that the introduction of an anti-conflict intervention increases the community's attention to conflicts.   Third, the significant negative effect of students' grade shows that senior students tend to perceive more conflicts than junior students.  Fourth, it can be seen that White and Asian-American students have more significant positive ratings on their school's atmosphere, while Black and Hispanic/Latino students have more negative views, especially for Black students based on the strongly negative  coefficient of this group. Such a finding may be due to the potential racial discrimination at school. Fifth, students living with both parents tend to have higher ratings on their school's atmosphere; this is probably because of the good impacts from their  family relationships. Sixth, it turns out that the more time students are disciplined, the more negative perception  they have toward their school's atmosphere. Lastly, the other variables including gender and whether the student was a return student from the previous year are not significant.

For comparison, we  also fit the SAR model (\ref{eq:SAR0})   with  $\bmW$ constructed based on another survey conducted at the end of the school year. To save space, the fitting results and their associated findings are provided in Table S.18 and Section S.10 of the supplementary material.   As recommended by an anonymous referee, we also present the results of using the 2SLS estimation by \citet{kelejian1998generalized} and the best 2SLS estimation by \citet{lee2003best} in Table S.19 of the supplementary material. 
Compared to the efficiency improved estimate $\hat{\tilde \bmtheta}$ in Tables \ref{tb:data1} and S.18, we find that both 2SLS estimates yield consistent results in terms of signs and significance.
Moreover, the efficiency improved estimates exhibit smaller standard errors and are notably closer to those of the QMLE.

\section{Conclusion}
\label{sec:conclusion}

This paper proposes a new quasi-score matching estimation approach for the SAR model. Based on our understanding, this article is the first to introduce score matching for spatially dependent data, unlike its prior use for independent and identically distributed data in machine learning (see, e.g., \citealp{hyvarinen2005estimation}). 
In addition, we propose an efficiency improved estimator that further approaches the efficiency of the QMLE. 
Both theoretical and numerical studies show that as $n$ is large, the efficiency differences among our two proposed estimators and the QMLE are small, while our estimators offer significantly simpler and faster computations. 
To analyze the asymptotic properties of these estimators under random weights matrix $\bmW$ and regressors $\bmX$, we develop a pioneering theoretical framework, including a new LLN and CLT. 
Both the quasi-score matching estimation approach and the proposed theoretical framework can broad applications in many other spatial econometrics  models (see, e.g., \citealp{lee2010estimation}, \citealp{zou2021network} and \citealp{wu2021inward}) for future research, in  particularly when both $\bmW$ and $\bmX$ are random.

  In addition to the asymptotic theory developed in this paper for random $\bmW$ and $\bmX$, as suggested by an anonymous reviewer, a promising direction for future research is to extend the theory of \citet{xu2015maximum} and \citet{wu2022applications} to allow for random $\bmW$. This would involve establishing near-epoch or spatial functional dependence properties for $y_i$ and related variables under model (\ref{eq:SAR0}), thereby generalizing, for example, Proposition 1 in \citet{xu2015maximum} or Proposition 4.3 in \citet{wu2022applications} to the random $\bmW$ setting. Section S.6 of the supplementary material also demonstrates that this generalization will facilitate the development of the preliminary conditions for the LLNs in Assumptions \ref{as:4} and \ref{as:4}$'$.
 We believe these efforts would strengthen the usefulness of  the  theoretical framework for random weights matrix $\bmW$ and regressors $\bmX$.

\section*{Acknowledgments}

We would like to thank the two anonymous referees, an Associate Editor and the Editor for their constructive comments, which helped us improve the quality of the paper. We also would like to thank Prof. Andrew Wood and A/Prof. Janice Scealy for introducing us to the field of score matching. The research of Xuan Liang and Tao Zou was supported by the assistance of computational resources provided by the Australian Government through the National Computational Infrastructure (NCI) under the ANU Startup Allocation Scheme.

\section*{Disclosure Statement}

The authors report there are no competing interests to declare.

\bibliographystyle{apalike}
\spacingset{1}

\bibliography{Bibliography}
\end{document}